\documentclass[11pt]{article}
\usepackage{latexsym}
\usepackage{amsmath}

 \csname
@addtoreset\endcsname{equation}{section}

\def\pe{\prime}
\def\3s{{s \choose 3}}
\def\4s{{s \choose 4}}
\def\5s{{s \choose 5}}
\def\6s{{s \choose 6}}
\def\12{\frac{1}{2}}
\def\fr{\frac}
\def\pr{\partial}
\def\prd{\partial \cdot}

\def\bec{\begin{center}}
\def\ec{\end{center}}
\def\a{\alpha}  

\def\b{\beta}

\def\d{\delta} 
\def\D{\Delta}
\def\e{\epsilon}

\def\vf{\varphi}

\def\l{\lambda}

\def\m{\mu}

\def\r{\rho}

\def\h{\eta}

\def\cB{{\cal B}}
\def\cY{{\cal Y}}
\def\cG{{\cal G}}

\def\g{\gamma}
\def\cJ{{\cal J}}
\def\cC{{\cal C}}
\def\cL{{\cal L}}
\def\cD{{\cal D}}
\def\cF{{\cal F}}

\def\cA{{\cal A}}
\def\cW{{\cal W}}
\def\cZ{{\cal Z}}

\def\cR{{\cal R}}
\def\cS{{\cal S}}

\def\cH{{\cal H}}

\def\cA{{\cal A}}

\newcommand{\eq}[1]{(\ref{#1})}

\def\be{\begin{equation}}
\def\ee{\end{equation}}
\def\bea{\begin{eqnarray}}
\def\eea{\end{eqnarray}}
\def\ba{\begin{array}}
\def\ea{\end{array}}

\def\scs#1{\section{\sc #1}}
\def\scss#1{\subsection{\sc #1}}

\def\dsll{\not {\! \pr}}
\def\psisl{\not {\! \! \psi}}

\def\cWsl{\not {\!\!\! \cal W}}
\def\e{\epsilon}
\def\esl{\not {\! \epsilon}}

\def\ssl{\not {\! \cal S}}
\def\xisl{\not {\! \xi}}
\def\esl{\not {\! \epsilon}}
\def\ssl{\not {\! \cal S}}
\def\nablasl{\not {\! \nabla}}
\def\Dsl{\not {\! \! \D}}

\def\ba{\begin{align}}
\def\ena{\end{align}}
\def\be{\begin{equation}}
\def\ee{\end{equation}}
\def\fr{\frac}

\def\a{\alpha}
\def\b{\beta}
\def\g{\gamma}

\def\d{\delta}
\def\D{\Delta}
\def\e{\epsilon}

\def\h{\eta}

\def\l{\lambda}

\def\m{\mu}

\def\r{\rho}

\def\vf{\varphi}

\def\cB{{\cal B}}

\def\cG{{\cal G}}
\def\cJ{{\cal J}}

\def\cC{{\cal C}}
\def\cL{{\cal L}}
\def\cD{{\cal D}}
\def\cF{{\cal F}}

\def\cA{{\cal A}}
\def\cW{{\cal W}}

\def\cY{{\cal Y}}
\def\cZ{{\cal Z}}

\def\cR{{\cal R}}
\def\cS{{\cal S}}

\def\cH{{\cal H}}

\def\cA{{\cal A}}

\def\dsll{\not {\! \pr}}

\def\psisl{\not {\! \! \psi}}

\def\esl{\not {\! \epsilon}}
\def\ssl{\not {\! \cal S}}

\def\xisl{\not {\! \xi}}
\def\xibsl{\not {\! \bar{\xi}}}
\def\esl{\not {\! \epsilon}}
\def\ssl{\not {\! \cal S}}

\def\nablasl{\not {\! \nabla}}
\def\psl{\not {\! \! p}}
\def\psll{\not { p}}

\def\gz0{\gamma^{0}}

\def\pe{\prime}
\def\eq{\equiv}

\def\pr{\partial}

\def\3s{{s \choose 3}}
\def\4s{{s \choose 4}}
\def\5s{{s \choose 5}}
\def\6s{{s \choose 6}}
\def\12{\frac{1}{2}}

\def\prd{\pr \cdot}

\begin{document}

\begin{flushright}
hep-th/yymmddd \vskip 8pt {\today}
\end{flushright}

\vspace{5pt}

\begin{center}


{\Large\sc Current Exchanges and Unconstrained Higher Spins}\\


\vspace{20pt}
{\sc D.~Francia${}^{a}$, J.~Mourad$^{b}$ and A.~Sagnotti$^{c}$}\\[15pt]

{${}^a$\it\sl\small Department of Fundamental Physics\\ Chalmers
University of Technology \\ S-412\ 96 \ G\"oteborg \ SWEDEN
\\ e-mail: {\small \it francia@chalmers.se}}\vspace{10pt}

{${}^b$\it\small AstroParticule et Cosmologie (APC) \footnote{Unit\'e mixte de Recherche du CNRS (UMR 7164)}\\
Universit\'e Paris VII - Campus Paris Rive Gauche \\
10, rue Alice Domon et Leonie Duquet \\ F-75205 Paris Cedex 13\
FRANCE
 \\ e-mail:
{\small \it mourad@apc.univ-paris7.fr}}\vspace{10pt}

{${}^c$\it\small
Scuola Normale Superiore and INFN\\
Piazza dei Cavalieri, 7\\I-56126 Pisa \ ITALY \\
e-mail: {\small \it sagnotti@sns.it}}

\vspace{10pt} {\sc\large Abstract}\end{center}

{The (Fang-)Fronsdal formulation for free fully symmetric (spinor-) tensors rests on ($\gamma$-)trace constraints on gauge fields and
parameters. When these are relaxed, glimpses of the underlying geometry emerge: the field equations extend to non-local expressions involving
the higher-spin curvatures, and with only a pair of additional fields an equivalent ``minimal'' local formulation is also possible. In this
paper we complete the discussion of the ``minimal'' formulation for fully symmetric (spinor-) tensors, constructing one-parameter families of
Lagrangians and extending them to $(A)dS$ backgrounds. We then turn on external currents, that in this setting are subject to conventional
conservation laws and, by a close scrutiny of current exchanges in the various formulations, we clarify the precise link between the local and
non-local versions of the theory. To this end, we first show the equivalence of the constrained and unconstrained local formulations, and then
identify a unique set of non-local Lagrangian equations which behave in the same fashion in current exchanges.}

 \setcounter{page}{1}

\pagebreak

\tableofcontents

\newpage

\scs{Introduction}\label{sec:Intro}

The (Fang-)Fronsdal formulation of free higher-spin dynamics
\cite{fronsdal} rests on trace (or $\gamma$-trace) constraints for
gauge fields and corresponding gauge parameters\footnote{The web
site http://www.ulb.ac.be/sciences/ptm/pmif/Solvay1proc.pdf contains
the Proceedings of the First Solvay Workshop on Higher-Spin Gauge
Theories \cite{solvay04}, with some contributions closely related to
the present work \cite{solvay1,bciv,sss} and many references to the
original literature.}. While these algebraic conditions do not
conflict with Lorentz covariance, it is quite natural to try and
forego them, aiming at formulations that are closer in spirit to the
familiar ones for low-spin fields. The issues are then how to do it
and what one gains from this extension.

A direct, if somewhat unconventional, way to forego the trace
constraints of \cite{fronsdal} is via a set of non-local equations
involving the higher-spin curvatures of de Wit and Freedman
\cite{dewfr}. This was first done in \cite{fs1} for the set of
higher-spin fields originally considered in \cite{fronsdal}, totally
symmetric tensors $\varphi_{\mu_1 \ldots \mu_s}$, that here will be
loosely referred to as spin-$s$ fields, and for corresponding
totally symmetric tensor-spinors $\psi_{\mu_1 \ldots \mu_n}$, that
here will be loosely referred to as spin-$(n+1/2)$ fields. In
exploring new directions for higher spins, it is customary to begin
by restricting the attention to this class of fields, although they
do not exhaust the available choices in $D>4$, since the resulting
analysis suffices to display some key features of the problem in a
relatively handy setting. The generalization to tensors of mixed
symmetry is then an important further step, and is also crucial for
establishing a proper link with String Theory. For the non-local
formulation this was first achieved to some extent in
\cite{nonsym1}, and the resulting properties were recently explored
in further detail in \cite{bb06}.

A less direct, but more conventional, way to forego the trace
constraints is via the introduction of compensators. This was first
achieved by Pashnev, Tsulaia, Buchbinder and others \cite{bpt},
using a BRST \cite{BRST} formulation inspired by String Field Theory
but adapted to the description of individual massless higher-spin
modes. This is to be contrasted with the conventional BRST
formulation of String Field Theory \cite{stringbrst}, that
associates to a given higher-spin field a whole family of other
fields belonging to lower Regge trajectories. The BRST formulation
of individual higher-spin fields of \cite{bpt} is rather
complicated, and makes use of ${\cal O}(s)$ fields to describe the
propagation of a single set of spin-$s$ modes. On the other hand, it
is possible to present ``compensator equations'' for individual
higher-spin fields that, while non-Lagrangian, are fully gauge
invariant without any trace constraints and involve only two fields.
For instance, in the bosonic case, aside from $\varphi_{\mu_1 \ldots
\mu_s}$ they involve a single compensator field $\alpha_{\mu_1
\ldots \mu_{s-3}}$, which first emerges for $s=3$. The relation with
free String Field Theory was clarified in \cite{fs2}, where the
precise link between the compensator equations for higher-spin
fields and the ``triplet'' systems \cite{oldtriplet,fs2,bonelli}
emerging from free String Field Theory in the low-tension limit was
displayed. In \cite{st} these equations were then related to the
results in \cite{bpt}, from which they can be recovered by a partial
gauge fixing, and were also extended to $(A)dS$ backgrounds.
Interestingly, the simplest flat-space equation in this set, for a
spin-3 field $\varphi_{\mu\nu\rho}$ and a scalar compensator
$\alpha$,
\be \Box \varphi_{\mu\nu\rho} - (\partial_\mu \partial \cdot
\varphi_{\nu\rho} + \ldots ) + (\partial_\mu \partial_\nu
\varphi^{\; \prime}_\rho + \ldots ) = 3\,
\partial_\mu\partial_\nu\partial_\rho \alpha \ , \ee
where, as in the rest of the present paper, ``primes'' denote
traces, was first considered by Schwinger long ago \cite{schw}
\footnote{We are grateful to G. Savvidy for calling Schwinger's
result to our attention.}. More recently, a ``minimal'' Lagrangian
formulation for the compensator equations was also obtained in
\cite{fs3}: rather than ${\cal O}(s)$ fields as in \cite{bpt}, it
involves only a Lagrange multiplier $\beta_{\mu_1 \ldots
\mu_{s-4}}$, that first emerges for spin $s=4$, aside from the basic
field $\varphi_{\mu_1 \ldots \mu_s}$ and the compensator
$\alpha_{\mu_1 \ldots \mu_{s-3}}$. In contrast with the non-local
case, the generalization of this ``minimal'' local formulation to
tensors of mixed symmetry is not known at the present time, although
its key features may be anticipated from the known constrained gauge
transformations.

While it is certainly interesting and instructive to explore them,
it is difficult to assess the relative virtues of different
formulations for free higher-spin fields before the systematics of
higher-spin interactions is further clarified. There have been many
attempts in this direction over the years \cite{interactionsold},
that have marked the history of the subject as a result of the
unexpected difficulties that were readily met, but this line of
approach deserves further efforts and is being further explored,
with the help of more powerful techniques, in recent times
\cite{interactionsrecent}. It is fair to say that we do not possess
yet a general understanding of higher-spin interactions and of the
underlying geometry, but we do have at our disposal two important
paradigmatic examples that are based on a well motivated algebraic
setting. These are the Vasiliev equations, in their four-dimensional
form based on spinor oscillators \cite{vas1} and in their more
recent $D$-dimensional form based on vector oscillators \cite{vas2}.
The Vasiliev equations are a set of first-order differential
constraints involving a one-form master field $A$, that encodes an
infinite family of $\varphi_{\mu_1 \ldots \mu_s}$ via corresponding
generalized vielbeins and spin connections, and a zero-form master
field $\Phi$, that collects their Weyl curvatures and covariant
derivatives, together with corresponding data for a scalar mode. Due
to the presence of the zero-form $\Phi$, the Vasiliev constraints
generalize the more conventional notion of free-differential algebra
\cite{fda} in a non-trivial fashion, and embody a description of
free higher-spin modes in an ``unfolded'' form, via infinitely many
auxiliary fields that subsume their local data. Both the
four-dimensional formulation of \cite{vas1} and the $D$-dimensional
formulation of \cite{vas2} are not Lagrangian, but while the former
is fully based on the constrained Fronsdal form of the free
dynamics, this is not quite true for the latter. More precisely, the
spinor oscillators build consistent interactions for the ``frame''
version of Fronsdal's formulation, which was first discussed in
\cite{lopvas}, and can be gauge-fixed to its ``metric'' version but
does not leave room for the compensator of \cite{fs2,st,fs3} or for
the wider gauge symmetry of \cite{bpt}. On the other hand, the field
equations of \cite{vas2}, as pointed out in \cite{sss}, allow a
non-dynamical ``off-shell'' variant devoid of trace constraints and,
if completed with a ``strong'' projection, turn into dynamical
equations that at the free level reduce precisely to a frame version
of the compensator equations of \cite{fs2,st,fs3}. The potential
open problems with the resulting interactions are still not fully
sorted out at the present time, and we refer the reader to
\cite{sss,bciv} for further details, but there are reasons to
believe that fully interacting equations can be defined in this
fashion, thus encoding naturally a gauge symmetry not subject to
Fronsdal's trace constraints.

To summarize, even with our current incomplete grasp of the
systematics of higher-spin interactions the unconstrained
formulation of free higher-spin fields presents a number of
attractive features. First, and most notably, it embodies a direct
link with higher-spin geometry, since its minimal, non-local form of
\cite{fs1} rests directly on the higher-spin curvatures of
\cite{dewfr} rather than on lower connections. In addition, it links
naturally to the BRST form of free String Field Theory, from which
it emerges, after a suitable truncation to the leading Regge
trajectory, in the low-tension limit \cite{fs2,st}. Finally, and
most importantly for the purposes of the present paper, it couples
to \emph{conserved} currents, to be contrasted with the partially
conserved currents of the Fang-Fronsdal formulation.

This paper is devoted to exploring some key features of the
``minimal'' unconstrained Lagrangian formalism of \cite{fs3} that
manifest themselves in the presence of external currents. However,
we begin in Section 2 by reconsidering and simplifying somewhat the
results of \cite{fs3} for bosons and fermions, that here are also
extended to the interesting cases of $(A)dS$ backgrounds.

While the extension of flat-space results to these more general backgrounds is an interesting and well-defined problem in its own right, this is
perhaps the place to spend, as in \cite{sss}, some words of caution against the naive identification of the spin-2 modes of the Vasiliev
equations with the gravitational field. The problem arises since, by virtue of their very field content, the Vasiliev equations are naturally an
effective description of the first Regge trajectory of the \emph{open} bosonic string, albeit in an unconventional and largely unexplored regime
where it would experience a gigantic collapse of all its massive modes to the massless level. It should be stressed that little is presently
known about this stringy regime, although the AdS/CFT correspondence, in the weak gauge-coupling limit, provides indirect arguments for it
\cite{adscft}. The Vasiliev equations also allow a Chan-Paton \cite{cp} extension to matrix-valued modes, precisely as is the case for the open
bosonic string. Hence, they generally involve not a single ``graviton'', but rather a whole multiplet of spin-2 modes, so that the ``minimal''
Vasiliev model, whose interest was clearly stressed in \cite{analysis} and whose first non-trivial ``cosmological'' solution was recently
presented in \cite{cosmo}, is somehow the counterpart in this context of the $O(1)$ open string discussed by Schwarz in his 1982 review
\cite{schwarzrev}, whose open sector contains indeed only even excitation levels. Should one then associate the singlet spin-2 Vasiliev mode
with gravity, that lies outside the open spectrum? Our present knowledge does not allow sharp statements to this effect, and caution is called
for. Indeed, while the distinction between open and closed spectra is strictly in place for tensile strings, in the largely unexplored
low-tension limit a mixing between string states at different levels, and in particular between the singlet part of the spin-2 open-string field
and the closed-string graviton, might take place, although a clear understanding of this important issue from a string vantage point is lacking
at present.

 In a similar spirit, a closed-string analogue of the
Vasiliev equations, which should rest on a more subtle algebraic
structure in order not to allow a Chan-Paton extension, is not
available at the present time, and the search for it might provide
important clues on low-tension string regimes. These are clearly
deep issues on which, unfortunately, we shall have little to say in
the present paper. Taking a close look at current exchanges,
however, in Sections 3 and 4 we shall be able to investigate in some
detail to which extent the available options for formulating the
free higher-spin dynamics lead to the same counting of degrees of
freedom. Thus, in Section 3 we shall establish the direct
equivalence between the constrained Fronsdal formulation of
\cite{fronsdal} and the unconstrained local formulation of
\cite{fs3}. On the other hand, the nonlocal counterpart of the
unconstrained free theory is not fully determined. Many possible
options for a non-local spin-$s$ Lagrangian equation exist and, as
we shall see, the simple choice made in \cite{fs1} for an
Einstein-like tensor \emph{does not} provide the direct counterpart
of the local Lagrangian equations of \cite{fs3}. In Section 4,
however, we shall identify a different, unique form of the non-local
Lagrangian field equations which behaves exactly like the local
formulations, thus arriving at a precise link between the local and
non-local unconstrained forms of free higher-spin gauge theory. Let
us stress that the issue here is the correct coupling to external
currents via the natural source term, $\varphi \cdot J$, and the
Lagrangian equations that couple correctly to external currents will
be more complicated than those proposed in \cite{fs1}. Still, in the
absence of external currents they can be turned into the non-local
non-Lagrangian equations of \cite{fs1},
\be \frac{1}{\Box^{\, p}} \, \prd  {\cal R}^{\, [p]}{}_{;\, \alpha_1 \cdots
\alpha_{2p+1}} \,  = \,  0 \label{oddcurv} \ee
for odd spins $s=2p+1$, and
\be \frac{1}{\Box^{\, p-1}} \ {\cal R}^{\, [p]}{}_{;\, \alpha_1 \cdots
\alpha_{2p}} \ =\ 0 \label{evencurv} \ee
for even spins $s=2p$. Section 5 contains our  Conclusions, while
the Appendix collects some useful results concerning our implicit
notation for symmetric tensors and our conventions.


\scs{Minimal Lagrangians for unconstrained higher
spins}\label{sec:min}

In this Section we review the construction of the minimal
unconstrained free Lagrangians for fully symmetric higher-spin
tensors and tensor-spinors in flat space presented in \cite{fs3}.
Our aim is to streamline both the derivation and the resulting
presentation, by properly stressing the role of a few gauge
invariant constructs playing the role of ``building blocks'' for the
dynamical quantities of interest. As we shall see, the resulting
simplified form of the Lagrangians proves quite helpful in extending
the previous results to the interesting cases of $(A)dS$ backgrounds.


\scss{Bosons}

The definition of fully gauge invariant kinetic operators is a very
convenient starting point for the construction of the minimal
bosonic Lagrangians, that we would like to reconsider and extend
here. For a rank-$s$ fully symmetric tensor, that in the index-free
notation of \cite{fs3} (see the Appendix for details and some
examples) can be simply denoted by $\varphi$, one can begin by
considering the Fronsdal operator
\be {\cF} \, = \, \Box \; \vf \, - \, \partial \, \partial \cdot \vf
\, + \, \partial^{\, 2} \vf^{\, \pe} \, , \label{fronsdop} \ee
where, here as elsewhere in this paper, a ``prime'' denotes a trace. Under the gauge transformation
\be
\delta \, \vf \, = \, \partial\, \Lambda \, ,
\label{transfgauge}
\ee
$\cF$ varies according to
\be \delta \, {\cF} \, = \, 3 \, \pr^{\, 3} \, \Lambda^{\, \pe} \, ,\ee
where, as anticipated, $\Lambda^{\, \prime}$ denotes the trace of the
gauge parameter $\Lambda$. From $\cF$ one can build the {\it fully} gauge
invariant tensor
\be \label{tensorA}
\cA \, = \, {\cF} \, - \, 3 \, \pr^{\, 3} \, \alpha \, ,
\ee
where the spin-$(s-3)$ \emph{compensator} $\a$ transforms as
\be \label{transalfa}
\delta \, \a \, = \, \Lambda^{\, \prime}
\ee
under the tensor gauge transformation in eq.~(\ref{transfgauge}),
and one can then show that $\cA$ satisfies the Bianchi identity
\be \prd \cA \, - \, \frac{1}{2} \, \pr \, \cA^{\, \prime} \, = \, -
\, \frac{3}{2} \, \pr^{\, 3} \, \left(\vf^{\, \prime \prime} \, - \,
4 \, \pr \cdot \a \, - \, \pr \, \a^{\, \prime} \right)\ .
\label{bianchibose} \ee

With the ``mostly-positive'' space-time signature that we shall use
throughout, the minimal bosonic Lagrangians of \cite{fs3} can thus
be conveniently recovered starting from
\be {\cL}_0 \, = \, \frac{1}{2} \, \vf \, \left( \cA \, - \,
\frac{1}{2} \, \eta \, \cA^{\; \prime} \right) \, , \ee
which, on account of (\ref{fronsdop}) and (\ref{bianchibose}),
varies as
\be \delta \cL_0 \, = \, \frac{3}{4} \ {s \choose 3} \, \Lambda^{\,
\prime} \prd \cA^{\, \prime} \, - \, 3 \ { s \choose 4} \, \prd \prd
\prd \Lambda \, \left[ \vf^{\, \prime\prime} \, - \, 4\, \prd \a \,
- \, \pr \, \a^{\, \pe} \right]\,  \ee
under the tensor gauge transformations (\ref{transfgauge}) and
(\ref{transalfa}). These terms can be compensated adding
\be {\cL}_1 \, = \, - \, \frac{3}{4} \ { s \choose 3 } \, \a \, \prd
\cA^{\, \prime} \, + \, 3 \, { s \choose 4 } \, \beta \, \left[
\vf^{\, \prime\prime} \, - \, 4 \, \prd \a \, - \, \pr \, \a^{\,
\prime} \right] \, , \ee
so that the end result can be presented in the rather compact form
\be \label{boselagr} {\cL} \, = \, \frac{1}{2} \, \vf \, \left( \cA
\, - \, \frac{1}{2} \ \eta \, \cA^{\; \prime} \right) \, - \,
\frac{3}{4} \ {s \choose 3 } \, \a\, \prd \cA^{\; \prime} \, + \, 3
\, { s \choose 4 } \, \beta \, \left[ \vf^{\; \prime\prime} \, - \,
4 \, \prd \a \, - \, \pr \, \a^{\, \pe} \right]\, , \ee
where, as in \cite{fs3}, the \emph{Lagrange multiplier} $\beta$
transforms as
\be \label{transbeta} \delta \beta \, = \, \prd \prd \prd \Lambda \,
.\ee

We would like to stress that working insofar as possible with the
gauge-invariant tensor $\cA$ has streamlined both the derivation and
the final form of the Lagrangian (\ref{boselagr}) with respect to
\cite{fs3}. One can actually do better defining, for the bosonic
case, three independent gauge-invariant tensors in terms of which
all dynamical quantities of interest can be expressed. The first
member of this set is the tensor $\cA$ introduced in
(\ref{tensorA}), the second,
\be \cC \, \equiv \, \vf^{\pe \pe} \, - \, 4 \, \prd \a \, - \, \pr \,
\a^{\, \pe} \, , \ee
defines the constraint, and finally the third,
\be \cB \, \equiv \, \b \, - \, \12 \, (\prd \prd \vf^{\, \pe} \, -
\, 2 \,  \Box \, \prd \a \, - \, \pr \, \prd \prd \a) \, , \ee
relates the Lagrange multiplier $\b$ to the single combination of
$\vf$ and $\a$ that possesses an identical gauge transformation.
Actually, the very definition of $\cB$ suggests that the Lagrangian
(\ref{boselagr}) is not the only possibility, and may be generalized
by turning the coefficient of $\cC$ into
\be \label{boselagrk} {\cL}_k \, = \, \frac{1}{2} \, \vf \, \left(
\cA \, - \, \frac{1}{2} \ \eta \, \cA^{\; \prime} \right) \, - \,
\frac{3}{4} \ {s \choose 3 } \, \a\, \prd \cA^{\; \prime} \, + \, 3
\, { s \choose 4 } \, [\b \, - \, k \, \cB] \, \cC\, , \ee
so that (\ref{boselagr}) corresponds to the particular choice $k \,
= \, 0$.

A more general analysis should involve additional quadratic terms
mixing $\cA$ and $\cC$ tensors, as well as terms quadratic in the
tensor $\cC$, thus exhausting all possible gauge-invariant terms
with at most two derivatives for the physical field $\vf$ and at
most four derivatives for the compensator $\a$. In particular, given
the identity
\be \cA^{\, \pe \pe} \, = \, 3\, \Box \, \cC \, + \, 3 \, \pr \,
\prd \, \cC \, + \, \pr^{\, 2} \, \cC^{\, \pe} \, , \ee
and since the only possibility to combine $\cA$ and $\cC$  without
increasing the number of derivatives in $\vf$ is via $\cA^{\, \pe
\pe} \, \cC$, it turns out that the only possibilities left are
quadratic terms in the tensor $\cC$,
\be \cC^{\, 2} \, , \hspace{.5cm} \cC \, \Box \, \cC \, ,
\hspace{.5cm} \prd \cC \, \prd \cC \,  , \hspace{.5cm} \cC^{\, \pe}
\, \prd \prd \cC \, ,\hspace{.5cm}  \ee
together with their traces. Nonetheless, all these terms would only
generate additional terms linear in $\cC$ in the equations for $\vf$
and $\a$ whose role would be irrelevant once $\cC$ is set to zero by
the field equation for $\b$.

We have thus arrived at a one-parameter family of gauge-invariant,
unconstrained Lagrangians, whose field equations can be nicely
expressed in terms of the three tensors $\cA, \, \cB$, and $\cC$,
and read
\be \label{boseeomk}
\begin{split}
E_{\vf} (k) & \equiv  \, \cA\, - \, \12 \, \h\, \cA^{\, \pe} \, +
\, \fr{1}{4} \, (1\, + \, k) \, \h \, \pr^{\, 2} \, \cC \,
+ \, (1\, - \, k) \, \h^{\, 2} \, \cB \, = \, 0 \, ,    \\
E_{\a} (k)  & \equiv \, -\, \fr{3}{2} {s \choose 3} \{ \prd \cA^{\, \pe} \,
- \, \fr{k\, + \, 1}{2} \, (\pr \, \Box  \, + \, \pr^{\, 2} \, \prd )\, \cC \,
+ (k\, - \, 1) \, (2\, \pr \, + \, \h \, \prd)\, \cB \} \, = \, 0 \, , \\
E_{\b} (k) & \equiv \, 3 \, { s \choose 4 }\, (1\,-\,k)\,\cC \, = \, 0 \, ,
\end{split}
\ee
where it should be noted that, in terms of the tensors
(\ref{boseeomk}), eq.~(\ref{boselagrk}) can be cast in the
particularly compact form
\be \cL_k \, = \, \12 \, \vf\, E_{\vf} (k) \, + \, \12 \, \a \,
E_{\a} (k) \, + \, \12 \, \b \, E_{\b} (k) \, . \ee

For generic values of $k$ the field equation for $\vf$ can be
reduced to the compensator form $\cA \, = \, 0$, while $\cB=0$
determines the Lagrange multiplier $\beta$ in terms of the other
fields, as was shown to be case in \cite{fs3} for $k=0$. One is
eventually left with the non-Lagrangian compensator equations of
\cite{fs2,st},
\be\label{nonlagr}
\begin{split}
&\cA \, \equiv \, {\cF} \, - \, 3 \, \pr^{\, 3} \, \alpha \, = \, 0 \, , \\
&\cC \, \equiv \, \vf^{\, \pe \pe} \, - \, 4 \, \prd \a \, - \, \pr
\, \a^{\, \pe} \, = \, 0 \, ,
\end{split}
\ee
and these can be finally gauge-fixed to the Fronsdal form $\cF = 0$
and to Fronsdal's double-trace constraint $\vf^{\, \pe \pe}=0$, making
use of the trace $\Lambda^{\, \prime}$ of the gauge parameter to
remove the compensator $\a$. This indicates that the unconstrained
Lagrangians (\ref{boselagrk}) provide a proper description of
higher-spin dynamics, just like Fronsdal's constrained formulation.

The value $k = -1$ appears particularly interesting, since in this
case the tensor $\cC$ disappears from the field equations for $\vf$
and $\a$. Finally, the choice $k = +1$ is somewhat degenerate, since
in this case $\b$ disappears from the Lagrangian, and this implies,
in its turn, that $\cB$ disappears from the field equations for
$\vf$ and $\a$. Whereas naively this could be seen as a
simplification, in this case one can only derive directly from the
field equations the condition
\be \Box \, \cC \, = \, 0 \, , \ee
which is not sufficient to remove the double trace.

In all cases, however, the divergence of $E_{\vf}(k)$ vanishes when
$\a$ and $\b$ are on-shell, as demanded by the Noether relation
\be \label{boseconscurr} \prd E_{\vf} (k)\, = \, \frac{1}{3\, {s \choose 3}} \,
\h \, E_{\a} (k) \,
- \, \frac{1}{4 \, {s \choose 4}} \,
\pr^{\, 3} \, E_{\b} (k) \, , \ee
that implies that external sources coupled to the dynamical field
$\vf$ are to be \emph{conserved}.

As anticipated, this presentation of the Lagrangians has the
virtue of leading rather directly to their $(A)dS$ deformations. These draw their
origin from the basic $(A)dS$ commutator for two covariant derivatives acting on a vector field $V$, that reads
\be \left[ \, \nabla_\mu \, , \nabla_\nu \, \right] \, V_\rho \, =
\, \frac{1}{L^2} \left(  g_{\nu\rho} V_\mu \, - \, g_{\mu\rho} V_\nu
\right) \, , \ee
where $L$ denotes the $(A)dS$ radius. For brevity, in the following we shall refer only to the $AdS$ case, since the corresponding results for
$dS$ backgrounds can be simply obtained by the formal continuation of $L$ to imaginary values.

In discussing higher-spin gauge fields in these curved backgrounds, it is convenient to begin by considering the deformed $AdS$ Fronsdal
operator \cite{fronadspap} (see also, for instance, \cite{st})
\be \cF_L \, = \, {\cal F} \, - \, \frac{1}{L^2} \, \left\{ \left[
(3-D-s)(2-s) - s \right]\, \varphi \ + \ 2 \, g \, \varphi^{'}
\right\} \, ,  \ee
where $D$ denotes the space-time dimension and
\be {\cal F} \, = \, \Box \, \varphi \, - \, \nabla \, \nabla \cdot
\varphi \, + \, \nabla^{\, 2} \, \vf^{\, \pe}\,  \ee
is the $AdS$-covariantized Fronsdal operator, that transforms according to
\be \delta {\cal F}_L \, = \, 3 \, \nabla^{\, 3} \Lambda^{\, \pe} \, - \, \frac{4}{L^2}\, g\,  \nabla \, \Lambda^{\, \pe}   \ee
under the gauge transformation
\be \delta \vf \, = \, \nabla \Lambda \ . \ee

Even in this more general $AdS$ setting the structure of the theory can be fully encoded in three gauge invariant tensors, $\cA_L$, $\cB_L$ and
$\cC_L$, that reduce to the previous expressions in the flat limit. Assuming for the auxiliary fields the straightforward generalisations of the
gauge transformations (\ref{transalfa}), (\ref{transbeta}), \be
\begin{split}
\d\, \a \,  &= \, \Lambda^{\, \pe} \, \\
\d \, \b \, &= \, \nabla \cdot \nabla \cdot \nabla \cdot \Lambda \, ,
\end{split}
\ee the $AdS$ generalizations of $\cA$ and $\cC$,
\begin{align}
\cA_L \, & = \, \cF_L \, - \, 3 \, \nabla^{\, 3} \a \, + \,
\frac{4}{L^2}\ g \, \nabla \,  \a \, , \\
\cC_L \, & = \, \vf^{\, \pe \pe} \, - \, 4 \, \nabla \cdot \a \, - \, \nabla \, \a^{\, \pe} \, ,
\end{align}
are rather simple, while the covariantization of $\cB$,
\be
\begin{split}
\cB_L \, & =  \, \b \, - \, \biggl\{\12 \, \nabla \cdot \nabla \cdot
\vf^{\, \pe} \, - \, \Box \, \nabla \cdot \a \, - \, \12 \, \nabla
\nabla
\cdot \nabla \cdot \a \\
            &  - \, \fr{1}{L^{\, 2}}\left( 2\, \nabla \a^{\, \pe} \,
+ \, 2\, g \, \nabla \cdot \a^{\, \pe} \, + \, \left[(s \, - \, 3)(5
\, - \, s \, - \, D)\right]\, \nabla \cdot \a \right) \biggr\}
\end{split}
\ee
is somewhat more tedious to obtain, since new terms appear in the
gauge variation of $\nabla \cdot \nabla \cdot \vf^{\, \pe}$ in this
curved background.

The starting point for the construction of the bosonic Lagrangians
is now
\be {\cL}_0 \, = \, \frac{e}{2} \, \vf \, \left( \cA_L \, - \,
\frac{1}{2} \ g\,  \cA_L^{\; \prime} \right) \, , \ee
where $e$ and $g$ denote the determinant of the vielbein and the $AdS$ metric.  Proceeding as in flat space and taking into account the deformed
Bianchi identity,
\be
\nabla \cdot {\cA}_L \, - \, \frac{1}{2} \, \nabla \,
{\cA}_L^{\, \pe} = - \, \frac{3}{2} \, \nabla^{\, 3} \, \cC_L
\, + \,  \frac{2}{L^{\, 2}} \, g \, \nabla \, \cC_L\, ,
\label{bianchiads}
\ee
one can finally arrive at the gauge invariant Lagrangians
\be \label{boselagrAdS}
{\cL} \, = \, \frac{e}{2} \, \vf \, \left( \cA_L \, - \,
\frac{1}{2} \ g\, \cA_L^{\, \pe} \right) \, - \, \frac{3e}{4} \ {s
\choose 3 } \, \a \, \nabla \cdot \cA_L^{\, \pe} \, + \, 3 \, e\, { s
\choose 4 } \, [\beta \, - \, \fr{4}{L^{\, 2}} \, \nabla \cdot \a] \, \cC_L \, .
\ee
that, as in the flat case, are special members of a whole family of
gauge invariant Lagrangians,
\be \label{boselagrAdSk} \cL_k \, = \, \frac{e}{2} \, \vf \, \left(
\cA_L \, - \, \frac{1}{2} \ g\, \cA_L^{\, \pe} \right) \, - \,
\frac{3e}{4} \ {s \choose 3 } \, \a \, \nabla \cdot \cA_L^{\, \pe}
\, + \, 3 \, e\, { s \choose 4 } \, \left[\b \, - \, \fr{4}{L^{\,
2}} \, \nabla \cdot \a \, - \, k \, \cB_L\right] \, \cC_L \, , \ee
defined introducing in $\cL$ the gauge-invariant coupling  $\cB_L \, \cC_L$. From these Lagrangians one can then derive the $AdS$ field
equations: for the fields $\vf$ and $\b$, they have the form of those obtained in flat case, aside from the natural substitutions
\be (\cA \, , \cB \, , \cC \, , \h \, , \pr) \, \longrightarrow \,
(\cA_L \, , \cB_L \, , \cC_L \, , g \, , \nabla) \, , \ee
while the field equation for $\a$ contains additional terms that
depend on the tensor $\cC_L$. The final result reads
\be \label{adsequations}
\begin{split}
E_{\vf\, L} (k) & \equiv  \, \cA_L\, - \, \12 \, g \, \cA^{\, \pe}_L \, + \, \fr{1}{4} \, (1\, + \, k) \, g \, \nabla^{\, 2} \, \cC_L \,
+ \, (1\, - \, k) \, g^{\, 2} \, \cB_L \, = \, 0 \, ,     \\
E_{\a\, L} (k)  & \equiv \, -\, \fr{3}{2} {s \choose 3} \left\{ \nabla \cdot \cA^{\, \pe}_L \, - \, \fr{k\, + \, 1}{2} \, \nabla \cdot
(\nabla^{\, 2}\, \cC_L) \, + (k\, - \, 1) \, (2\, \nabla \, + \, g \, \nabla \cdot)\, \cB_L \, - \,
\fr{4}{L^{\, 2}} \nabla \, \cC_L\right\} \, = \, 0 \, , \\
E_{\b\, L} (k) & \equiv \, 3 \, { s \choose 4 }\, (1\,-\,k)\,\cC_L \, = \, 0 \, .
\end{split}
\ee
More explicitly,  in the case $k \, = \, 0$ the field equation for
$\vf$  takes the form
\be
\begin{split}
E_{\vf\, L} \, \equiv & \, \cA_L \, - \, \frac{1}{2} \, g \, \left\{ \cA_L^{\, \pe} \, - \, \frac{1}{2} \nabla^{\, 2} \, \left( \vf^{\, \pe \pe}
- 4\, \nabla \cdot \alpha \, -
\, \nabla \, \a^{\, \pe} \right) \right\}  \\
& + \, g^2 \, \left\{ \b \, +\, \frac{1}{2} \, \nabla \nabla
\cdot \nabla \cdot \a \, +\, \Box \nabla \cdot \a \, - \,
\frac{1}{2} \ \nabla \cdot \nabla \cdot \vf^{\, \pe} \, + \,
\frac{2}{L^2} \ \nabla \alpha^{\, \pe} \, \right. \\
&+\left. \, \frac{2}{L^2} \ g \, \nabla \cdot \a^{\, \pe} \, + \,
\frac{1}{L^2} \, \left[ (s-3) (5 - s - D) + 4 \right]\, \nabla \cdot
\a \right\} \, .
\end{split}
\ee

Current conservation is guaranteed in this case by the $AdS$ generalization of eq.~(\ref{boseconscurr}),
\be \label{boseconscurrAdS} \nabla \cdot E_{\vf\, L} (k)\, = \, \frac{1}{3\, {s \choose 3}} \, g \, E_{\a\, L} (k) \, - \, \frac{1}{4 \, {s
\choose 4}} \, \nabla^{\, 3} \, E_{\b\, L} (k) \, ,
\ee%
a result that reflects the Noether identity signalling the gauge
invariance of the action, which indeed implies that
\be \int \, d^Dx \ \left[- \ s \ \Lambda \ \nabla \, \cdot \,
\frac{\delta {\cal L}_k}{\delta \vf} \, + \, \nabla \cdot \nabla
\cdot \nabla \cdot \Lambda \ \frac{\delta {\cal L}_k}{\delta \b} \,
+ \, \Lambda^{\; \prime} \, \frac{\delta {\cal L}_k}{\delta \alpha}
\right] \, = \, 0 \ . \ee


\scss{Fermions}


The fermionic case is more involved but does not add substantial novelties with respect to what we have seen for bosons. As in the preceding
Subsection, we can now begin by simplifying and generalizing the construction of gauge invariant Lagrangians for unconstrained free fields of
\cite{fs3}. In doing so, we shall be able to stress the key role of a few gauge-invariant blocks, a step that will prove again quite convenient
when constructing the $AdS$ deformation of the flat-space results. Let us therefore begin by recalling the definition of the Fang-Fronsdal
operator for a totally symmetric rank-$n$ tensor-spinor $\psi$ \cite{fronsdal},
\be {\cS} \, = \, i\, \left(\dsll \, \psi \, - \, \pr \, \psisl
\right) \, , \ee
where $\psisl$ denotes the $\gamma$-trace of the gauge field.
Under the gauge transformation
\be \delta \, \psi \, = \, \partial\, \e \, , \label{tensorgauge}
\ee
$\cS$ varies according to
\be \delta \, {\cS} \, = \, -  \, 2 \, i\, \pr^{\, 2} \esl \, , \ee
where $\not {\! \epsilon}$ denotes the $\gamma$-trace of the
gauge parameter $\e$. In analogy with what we have seen for bosons,
one can build from $\cS$ the {\it fully} gauge invariant operator
\be {\cW} \, \equiv \, S \, + \, 2 \, i\, \pr^2 \xi \, , \ee
where the rank-$(n-2)$ \emph{compensator} $\xi$ transforms as
\be \delta \, \xi \, = \, \esl \, \label{spinorgauge} \ee
under the gauge transformation (\ref{tensorgauge}).

The Bianchi identity for $\cW$,
\be \prd \cW \, - \, \frac{1}{2} \ \pr \, \cW^{\, \pe} \, - \,
\frac{1}{2} \, \dsll \, \, {\cWsl}  \, = \, i\, \pr^{\, 2} \left[ \,
\psisl^{\, \pe} \, - \, 2 \, \prd \xi \, - \, \pr \, \xi^{\, \pe} -
\not{\!  \pr} \xisl \, \right] \, , \label{bianchifermi} \ee
leads naturally to a second gauge-invariant tensor-spinor,
\be \cZ \, \equiv \, i\, \bigl\{ \psisl^{\, \prime} \, - \, 2 \,
\prd \xi \, - \, \pr \, \xi^{\, \prime} - \not{\!  \pr} \xisl\bigr\}
\, , \ee
directly related to the triple $\g$-trace constraint on the
fermionic gauge field $\psi$, that is absent in the Fang-Fronsdal
formulation. The minimal flat-space Lagrangians of \cite{fs3} can
then be recovered starting from the trial Lagrangians
\be \cL_0 \, = \, \12 \, \bar{\psi} \, \left( \cW \, - \,
\frac{1}{2} \, \eta \, \cW^{\, \pe} \, - \, \frac{1}{2} \, \g \,
\cWsl \right) \ + \ h.c. \, , \ee
and compensating the remainders in their gauge transformations with
new terms involving the field $\xi$ and the tensor $\cZ$. The
complete Lagrangians are finally
\be \label{fermilagr}
\begin{split}
\cL = & \12 \, \bar{\psi} \, \left( \cW \,- \, \frac{1}{2} \, \g \cWsl\, - \,
\frac{1}{2} \, \eta \, \cW^{\, \pe}
\right) \, - \, \frac{3}{4} \, {n \choose 3} \,  \xibsl \, \prd
\cW^{\, \pe} \\
& +\, \frac{1}{2} \, {n \choose 2} \, \bar{\xi} \, \prd \cWsl \, + \,
\frac{3}{2} {n \choose 3} \bar{\lambda} \, \cZ \,  + \ h.c. \, ,
\end{split}
\ee
where, as in \cite{fs3}, we have introduced a  \emph{Lagrange
multiplier} $\lambda$, whose gauge transformation is determined to
be
\be \delta \lambda \, = \, \prd \prd \epsilon \ee
in order for $\cL$ to be gauge invariant. As in the bosonic case, it
is useful to introduce an additional gauge-invariant tensor, now
involving $\lambda$,
\be \cY \, \equiv \, i \, \left[ \lambda \, - \, \12\, (\prd \psi^{\, \pe}
\, - \, \Box \, \xisl \, - \, \pr \, \prd \, \xisl)\right] \, . \ee
since only this gauge-invariant combination can actually enter the
field equations.

The identification of the $\cY$ tensor suggests again that
(\ref{fermilagr}) is but a member of a family of Lagrangians
depending on a parameter $k$,
\be \label{fermilagrk}
\begin{split}
\cL \, =& \12 \, \bar{\psi} \, \left( \cW \, - \,
\frac{1}{2} \, \eta \, \cW^{\, \pe} \, - \, \frac{1}{2} \, \g \cWsl
\right) \, - \, \frac{3}{4} \, {n \choose 3} \, \xibsl \, \prd
\cW^{\, \pe}  \\
& + \frac{1}{2} \, {n \choose 2} \, \bar{\xi} \, \prd \cWsl \, + \,
\frac{3}{2} \, {n \choose 3}\, (\bar{\lambda} \, - \, i\, k \, \bar{\cY}) \, \cZ
\,  + \ h.c. \, ,
\end{split}
\ee
where the previous case of eq.~(\ref{fermilagr}) corresponds to the
choice $k = 0$. Other possible  quadratic terms in $\cZ$
are excluded as irrelevant, for the same reasons as
in the bosonic case. The field equations following from
(\ref{fermilagrk}) are then
\be \label{fermieomk}
\begin{split}
E_{\bar{\psi}} (k) & \equiv  \, \cW \, - \, \frac{1}{2} \, \g \cWsl \,
- \, \12 \, \eta \, \cW^{\, \pe} \,\, - \, \fr{1}{4} \, (1\, + \, k) \, \h \, \pr \, \cZ \, - \,
\12 \, (1\, - \, k) \, \h \, \g \, \cY \, = \, 0 \ ,    \\
E_{\bar{\xi}} (k)  & \equiv \,  {n \choose 2} \biggl\{ \prd
\cWsl \, + \, \12 \, \g \, \prd \, \cW^{\, \pe}
 + \, (1\, - \, k) \, \bigl(\pr \, + \, \12  \g \dsll \, + \, \h \, \prd\bigr)\, \cY  \\
& \ \ \ \ \ \ \ \ \ \ \ \ \ \ +\, \fr{1 \, + \, k}{4} \, \g \, (\Box \, +
\, \pr\,  \prd )\, \cZ\biggr\} \,
= \, 0 \ , \\
E_{\bar{\lambda}} (k) & \equiv \, \fr{3}{2} \, {n \choose 3}\, (1\, - \,
k)\, \cZ \, = \, 0 \ .
\end{split}
\ee
As in the bosonic case, they can be reduced in general to the
compensator equations of \cite{fs2,st}
\be
\begin{split}
& {\cW} \, \equiv \, S \, + \, 2 \, i\, \pr^{\, 2} \,  \xi \ = \ 0 \ , \\
& \cZ \, \equiv \, i\, \bigl\{ \psisl^{\, \prime} \, - \, 2 \, \prd
\xi \, - \, \pr \, \xi^{\, \prime} - \not{\!  \pr} \xisl\bigr\} \ = \ 0
\, ,
\end{split}
\ee
and eventually to the Fang-Fronsdal form upon partial gauge fixing.

In complete analogy with the bosonic case, when $\bar{\psi}$ is
coupled to a \emph{conserved} external source the consistency of the
system is guaranteed by the field equations for the two auxiliary
fields $\bar{\xi}$ and $\bar{\lambda}$, since
\be \label{fermiconscurrk} \prd E_{\bar{\psi}} (k)\, = \,
\frac{1}{3\, {n \choose 3}} \, \pr^{\, 2} \, E_{\bar{\lambda}} (k)
 \, - \, \fr{1}{2 \, {n \choose 2}}\, \g \,
\, E_{\bar{\xi}} (k) \, . \ee
Moreover, in terms of the tensors defined in (\ref{fermieomk}), the
Lagrangian finally takes the simple form
\be \cL_k \, = \, \12 \, \bar{\psi} \, E_{\bar{\psi}} (k)  \, + \,
\12 \, \bar{\xi} \, E_{\bar{\xi}} (k) \, + \, \12 \, \bar{\lambda} \,
E_{\bar{\lambda}} (k) \, + \, h. c. \ . \ee

Generalizing these expressions to an $AdS$ background entails a few complications. First of all, the gauge transformation of $\psi$ acquires an
additional contribution \cite{fangfrads}, a phenomenon that has a well-known counterpart in gauged supergravity, so that now
\be \delta \psi \, = \, \nabla \epsilon \, + \, \frac{1}{2\, L}\ \gamma
\, \epsilon \, . \label{fermiads} \ee
Hence, with our ``mostly plus'' convention for the space-time
signature,
\be \delta \bar{\psi} \, = \, \nabla \bar{\epsilon} \, - \,
\frac{1}{2\, L} \, \bar{\epsilon} \, \gamma \, . \ee
The fermionic gauge transformation lends naturally to define a
modified covariant derivative,
\be \label{trueder}
\D \, \equiv \, \nabla \, + \, \frac{1}{2\, L} \g \, ,
\ee
in terms of which the basic relations take a simpler form. The starting point is again the covariantized Fang-Fronsdal operator, that in an
$AdS$ background is also modified by the addition of mass-like terms and reads
\be \cS_L \, = \, i\, ( \Dsl \, \psi \, - \, \D \, \psisl) \, + \,
\fr{i}{L} \, (n \, - \, 2) \, \psi \, + \, \fr{i}{L} \, \g \, \psisl
\, , \ee
One can now show that $\cS_L$ transforms as
\be \d \, \cS_L \, = \, - \, i \, \left(\D^{\, 2} \esl \, - \,
\fr{2}{L} \, \g \, \D \esl\right)\, , \ee
under the gauge transformation and satisfies the Bianchi identity
\be \D \cdot \cS_L \, - \, \12 \, \Dsl \, \ssl_L \,-\,  \12 \, \D \,
 \cS_L^{\, \pe}  \, = \,  \frac{1}{2\,L} \, (n \, - \, 1) \,  \ssl_L \,
- \, \frac{1}{2\,L} \, \g \, \cS_L^{\, \pe} \, + \, \fr{i}{2} \,
\left(2 \, \D^{\, 2} \,  - \, \fr{2}{L} \, \g \, \D\right) \psisl^{\, \pe}
\,
 , \label{bianchiD}
\ee
a simpler expression than the corresponding one in terms of the more
conventional derivative $\nabla$,
\be
\begin{split}
\nabla \cdot {\cal S}_L \, - \, \12 \, \nablasl \, \ssl_L\, - \,
\frac{1}{2} \, \nabla \, \cS_L^{\, \pe}  = &- \,\frac{1}{4L} \,
\gamma \, {\cal S}^{\, \prime}_L \ + \ \frac{1}{4L} \left[ (D
\, - \, 2)\, +\, 2\; (n\, - \, 1) \right] \, \ssl_L  \\
&+ \, \frac{i}{2} \ \left(  2\,  \nabla^2 \, - \, \frac{1}{L}\ \gamma \, \nabla
\, - \, \frac{3}{2\;L^2}\ g \right) \, \psisl^{\, \pe} \, ,
\end{split}
\ee
where we are correcting a couple of misprints present in eq.~(5.37)
of \cite{st}. One may notice, in particular, that the space-time
dimension $D$ never appears explicitly in eq.~(\ref{bianchiD}).

As in the bosonic case, one is led to define the deformed
gauge-invariant structures
\be
\begin{split}
\cW_L \, &= \, \cS_L \, + \, 2 \,  i \, \D^{\, 2} \, \xi \, - \, \fr{2\, i}{L} \, \g \, \D \, \xi \, , \\
\cY_L \, &=  i\, \left\{\lambda  -  \12  (\D\cdot \psi^{\, \pe}  -  \Box
\xisl - \D  \, \D \cdot  \xisl \,  - \fr{1}{L^{\, 2}} [(n  -  3) (5
-  n  -  D)  \xisl
\, + \, 2 \,  g  \xisl^{\, \pe}] \right\} \, , \\
\cZ_L  \, &= i\, \left\{\psisl^{\, \pe} \, - \, 2 \, \D \cdot \xi \,
- (\, \D \, - \, \fr{1}{L} \, \g) \, \xi^{\, \pe} \, - \, (\Dsl \, + \, \fr{1}{L} \,
(n \, - \, 3))\, \xisl \,
\right\} \, ,
\end{split}
\ee
where in particular the $\cY_L$ tensor is gauge-invariant provided
the field $\lambda$ transforms according to
\be \d \, \lambda \, = \, \D \cdot  \D \cdot  \e \, . \ee

The Bianchi identity satisfied by $\cW_L$,
\be \D \cdot \cW_L \, - \, \12 \, \Dsl \, \cWsl_L \,- \,  \12 \, \D
\,
 \cW_L^{\, \pe}  \, = \,  \frac{1}{2\,L} \, (n \, - \, 1) \,  \cWsl_L \,
 - \, \frac{1}{2\,L} \, \g \, \cW_L^{\, \pe} \, + \, \fr{1}{2} \,
 \left(2 \, \D^{\, 2} \, - \, \fr{2}{L} \, \g \, \D\right) \cZ_L \, ,
\ee
or, what is equivalent, the divergence of the kinetic operator
constructed from it,
\be
\begin{split}
\D \cdot \left\{ \cW_L \, - \, \12 \, \g \, \cWsl_L\, - \, \12 \, g \,
 \cW_L^{\, \pe}  \right\} \, = \, - \, \12 \, \g \, \D \cdot \, \cWsl_L\,
 - \, \12 \, g \, \D \cdot \, \cW_L^{\, \pe} \, + \, \12 \,
 \left(2 \, \D^{\, 2} \, - \, \fr{2}{L} \, \g \, \D\right) \cZ_L \, ,
\end{split}
\ee
is then the main ingredient to build the one-parameter family of
gauge-invariant Lagrangians
\be
\begin{split}
\cL \, = \, & \fr{e}{2} \, \bar{\psi} \, \left( \cW_L \, - \, \12 \, \g
\cWsl_L \, - \,
\12 \, g \cW^{\, \pe}_L \right) \, - \, \frac{3e}{4} \, {n \choose 3} \,  \bar{\xisl}
\, \D \cdot \cW^{\, \pe}_L \\
& + \, \frac{e}{2} \, {n \choose 2} \,
\bar{\xi} \, \D \cdot \cWsl_L \,
  + \, \fr{3\, e}{2} \, {n \choose 3} \left(\bar{\lambda} \, - \, \fr{2}{L} \, \D \cdot \bar{\xi} \,-
  \, i \,  k \, \cY_L\right) \, \cZ_L \, +
  \, h.c. \, .
\end{split}
\ee From these one can finally derive the $AdS$ field equations,
\be
\begin{split}
E_{\bar{\psi} \, L} (k) & \equiv  \, \cW_L \, - \, \frac{1}{2} \, \g \cWsl_L \, - \, \12 \, g \cW^{\, \pe}_L \,
\, - \, \fr{1}{4} \, (1\, + \, k) \, g \, \D \, \cZ_L \, - \,
\12 \, (1\, - \, k) \, \g \, g \, \cY_L \, = \, 0 \, , \\
E_{\bar{\xi} \, L} (k)  & \equiv \,  {n \choose 2} \left\{\D \cdot \,
\cWsl_L \, + \, \12 \, \g \, \D \cdot \, \cW^{\, \pe}_L + \, (1\, - \,k)
\, \left(\D \, \cY_L \,+\, \fr{1}{L} \, \D \cdot \, (\g \, \cY_L) \right) \right. \\
& \left. \ \ \ \ +\, \frac{2}{L} \, \D \, \cZ_L \, + \, \fr{1 \, + \, k}{4} \, \g \,  \D \cdot (\D\,
\cZ_L) \right\} \,
= \, 0 \, , \\
E_{\bar{\lambda}\, L} (k) & \equiv \, \fr{3}{2} \, {n \choose 3}\, (1\, -
\, k)\, \cZ_L \, = \, 0 \, .
\end{split}
\ee
The conservation on an external current coupled to $\psi$ is finally
implied by the analog of (\ref{fermiconscurrk}),
\be \D \cdot E_{\bar{\psi} \, L} (k) \, = \, \frac{1}{6\, {n \choose
3}} \,
 \, \D^{\, 2} \, E_{\bar{\l}\, L} (k)
 \, - \, \fr{1}{2 \, {n \choose 2}}\, \g \,
\, E_{\bar{\xi}\, L} (k) \, . \ee
%


\scs{Current exchanges in the local formulations}


In this Section we consider the response of unconstrained
higher-spin gauge fields to external currents, focussing in
particular on the current exchange. This is a convenient device to
compare different formulations in the simplest possible setting, and
the ensuing analysis is indeed rather rewarding. As we shall see,
the constrained Fronsdal formulation agrees directly, in this
respect, with the minimal unconstrained form discussed in the
previous Section, but not with the non-local geometric formulation
of \cite{fs1,fs2}. The lack of direct agreement between the source
couplings in the local and non-local forms of the theory, however,
is an interesting fact that can be turned to our own advantage: it
determines a unique form for the non-local theory, selecting a
specific form for the corresponding Lagrangian.

In the absence of sources, as stressed in \cite{fs2}, the iterative
procedure of \cite{fs1} builds a sequence of pseudo-differential
operators
\be {\cal F}^{(n+1)} \ = \ {\cal F}^{(n)} \ + \ \frac{1}{(n+1) (2 n
+ 1)} \ \frac{\partial^{\;2}}{\Box} \, {{\cal F}^{(n)}}\;' \ - \
\frac{1}{n+1} \ \frac{\partial}{\Box} \
\partial \cdot  {\cal F}^{(n)} \ , \label{recursion}
\ee
turning the first of the compensator equations (\ref{nonlagr}) into
a sequence of non-local equations,
\be {\cal F}^{(n)} \ = \ (2 \, n \ + \ 1) \ \frac{\pr^{\, 2
n+1}}{\Box^{n-1}} \, \alpha^{[n-1]} \ . \label{iter} \ee
Eventually one thus arrives at an irreducible form involving the
gauge field $\varphi$ alone, that is to be expressible in terms of
higher-spin curvatures. This gauge invariant form, however, is
clearly not unique. For instance, after $\alpha$ has disappeared,
further iterations of (\ref{recursion}) will produce additional
gauge invariant equations. However, as we shall see in the next
Section, external sources select a unique form of the non-local
equations, that should be regarded as the proper counterpart of the
Lagrangian equations of the previous Section, although they will be
somewhat more complicated than the Lagrangian equations of
\cite{fs1}. We can now begin to investigate these issues in some
detail for the local formulation, in order to show the equivalence
of its constrained and unconstrained forms.

\scss{Bose fields in flat space}

In order to motivate our procedure, let us begin by rephrasing a simple
and familiar example, the comparison between the light-cone and
covariant forms of the Maxwell theory. In the former, one has the $D-2$
transverse components $A_i$ of the gauge field, that in momentum space
couple to an external current $j_i$ according to
\be p^2 \, A_i \ = \ j_i \ . \ee
The current-current interaction in this physical gauge is thus sized
by the product of two purely transverse currents, since
\be p^2 \, j_i A_i\ = \ j_i \, j_i \ . \label{lcgs1} \ee
In a similar fashion, in the covariant formulation one would start
from the Maxwell equation for the full vector potential,
\be \left( p^2 \, \eta_{\mu\nu} \ - \ p_\mu \, p_\nu \right)  A^\nu
\ = \ J_\mu \ ,\ee
where, however, consistency demands that the current be
\emph{conserved}. As a result, in this case the current-current
interaction is sized by the full scalar product of a pair of
covariant \emph{conserved} currents, since
\be p^2 \, J^\mu \, A_\mu \ = \ J^\mu \, J_\mu \ . \label{covs1} \ee
Incidentally, here one faces a cute, albeit well-known, fact:
folding the kinetic operator into a conserved current has made it
possible to effectively recover the propagator without a
gauge-fixing procedure. The issue is now to proceed to the
(singular) on-shell limit, in order to compare the number of degrees
of freedom exchanged in the two formulations. These are encoded in the
Euclidean product $j_i \, j_i$ for eq.~(\ref{lcgs1}) and in the
Lorentzian product $J_\mu \, J^\mu$ for eq.~(\ref{covs1}), which,
differently from the propagators, are well defined on-shell.

An arbitrary on-shell current $J_\mu(p)$ can be made transverse upon
multiplication by the projector
\be \Pi_{\mu\nu} \ = \ \eta_{\mu\nu} \ - \ p_\mu \bar{p}_\nu \ - \
p_\nu \bar{p}_\mu \ , \ee
where $p$ is the exchanged on-shell momentum, so that $p^2=0$, and $\bar{p}$
is a second vector such that $\bar{p}^2=0$ and $p \cdot \bar{p} =
1$. Notice that the projector $\Pi$ satisfies the conditions
\be \Pi_{\mu\nu}\, \Pi^{\nu\rho}\ =\ {\Pi_{\mu}}^{\rho}\ , \quad
\eta_{\mu\nu} \, \Pi^{\mu\nu}\ = \ D-2 \ ,\quad  \Pi_{\mu\nu}\,
p^\nu\ =\ 0 \ . \ee
The same number of polarizations is thus exchanged in both cases: in
the former one has directly $(D-2)$ of them in $D$ space-time
dimensions, while the latter involves $J^\mu \,\Pi_{\mu\nu}\,
J^\nu$, so that the trace of the projector $\Pi_{\mu\nu}$, again
equal to $D-2$, leads to the same result even starting with full
$D$-dimensional covariant currents.

One can repeat almost verbatim the exercise for a spin-2 field. For
all dimensions $D\geq 3$, its degrees of freedom in the light-cone
exchange fill the $\frac{D(D-3)}{2}$ independent components a
symmetric traceless tensor in $(D-2)$ dimensions, while now the
covariant Lagrangian equation is
\be p^2 \, h_{\mu\nu} \ - \ p_\mu \, p \cdot h_\nu \ - \ p_\nu \, p
\cdot h_\mu \ + \ p_\mu \, p_\nu \, h^{\; \prime} \ - \
\eta_{\mu\nu} \, \left( p^2 \, h^{\; \prime} \ - \ p \cdot p \cdot
h\right) \ = \ J_{\mu\nu}(p) \ . \ee
Combining this equation with its trace then gives
\be p^2 \, h_{\mu\nu} \ - \ p_\mu \, p \cdot h_\nu \ - \ p_\nu \, p
\cdot h_\mu \ + \ p_\mu \, p_\nu \, h^{\; \prime} \ = \ J_{\mu\nu} \
- \ \frac{\eta_{\mu\nu}}{D-2} \ J^{\; \prime} \, \label{spin2} \ee
and finally folding it, as above, with the conserved current
$J_{\mu\nu}(p)$,
\be p^2 J^{\mu\nu}\, h_{\mu\nu} \ = \  J^{\mu\nu}\, J_{\mu\nu} \ - \
\frac{1}{D-2} \ \left(J^{\;\prime}\right)^2 \ , \ee
or equivalently
\be p^2 J^{\mu\nu}\, h_{\mu\nu} \ = \  \left( J_{\mu\nu}\ - \
\frac{1}{D-2} \ \Pi_{\mu\nu}\, J^{\;\prime}\right)^2 \ , \ee
where the expression within brackets is the traceless and transverse
projection of $J$. In deriving this result we have taken into
account that, as in the spin-one case, the conservation of
$J_{\mu\nu}$ forces the projector $\Pi$ into the expression. The end
conclusion is, therefore, that the degrees of freedom interchanged
in the covariant formulation fill a traceless symmetric matrix in
$D-2$ dimensions, just like their light-cone counterparts.

The previous discussion gives us the flavor of the general case,
although it inevitably leaves out its key subtleties. Thus, in $D$
dimensions the degrees of freedom carried by a massless fully
symmetric tensor $\varphi_{\mu_1 \ldots \mu_s}$ should fill an
irreducible representation of the little group $SO(D-2)$
corresponding to a \emph{traceless} symmetric tensor $j_{a_1\dots
a_s}$. Standard properties of Young tableaux determine the dimension
of a traceful rank-$s$ symmetric tensor of $SO(D)$,
\be P(D,s)={(D+s-1)!\over (D-1)!\, s!}\ , \ee
while for a traceless symmetric tensor of the same rank in $D-2$
dimensions the corresponding number is
\be P(D-2,s) \ - \ P(D-2,s-2) \ = \ \frac{(D+2s-4) (D+s-5)!}{ (D-4)!
\, s!} \ . \ee
Only for $s=1,2$, however, can this result be also expressed as
$P(D,s)-2P(D,s-1)$, the relation suggested by the two simple
examples above. In both these cases, one can indeed first subtract
$P(D,s-1)$ degrees of freedom to account for the Lorentz (or de
Donder) gauge conditions, one for $s=1$ and a full vector for $s=2$,
and then again the same number, $P(D,s-1)$, to account for a second,
on-shell, gauge transformation that preserves the first. This is at
the origin of the simple description of the $s=1$ (Maxwell) and
$s=2$ (Einstein) cases by unconstrained fields with a local gauge
invariance, while the discrepancy for $s>2$ reflects the novel
features of higher-spin fields, the need for Fronsdal's trace
conditions as in \cite{fronsdal} or for the compensators of the
previous Section.

Returning to our main task we would like to discuss, in a
Lorentz-covariant formalism and for an arbitrary tensor
$\varphi_{\mu_1 \ldots \mu_s}$, the current exchange for a pair of
totally symmetric currents $J_{\mu_1 \ldots \mu_s}$. As anticipated,
the light-cone construction forces the result to equal $j_{a_1\dots
a_s}\, j^{a_1\dots a_s}$ on shell, and the issue at stake is
whether, in the various available formulations, the basic equality
\be J_{\mu_1 \ldots \mu_s} \ {\cal P}^{\mu_1 \ldots \mu_s ; \nu_1
\ldots \nu_s} \ J_{\nu_1 \ldots \nu_s} \ = \ j_{a_1\dots a_s}\
j^{a_1\dots a_s} \ , \label{compare} \ee
holds, with ${\cal P}$ the spin-$s$ analogue of the tensor defined
by the \emph{r.h.s.} of eq.~(\ref{spin2}). In the following we shall
describe how this identity, that guarantees that only physical
degrees of freedom are exchanged, can be recovered in the local,
constrained or unconstrained, formulations of higher-spin gauge
fields. In Fronsdal's construction, where both the double trace of
the current $J$ and the traceless part of its divergence vanish,
this result was presented in \cite{fronsdal}, and the derivation is
repeated here for completeness, while simply extending it to $D$
dimensions.

We can now build the \emph{l.h.s.} of eq.~(\ref{compare}). To this
end, let us begin by noticing that, given a generic totally
symmetric current $J$, its projection onto a traceless symmetric
tensor can be attained via the sum
\be T_s J\ =\ \sum_{0}^{N}\rho_n(D,s)\, \eta^{n}\, J^{[n]},
\label{traceless} \ee
where $J^{[n]}$ denotes the $n^{th}$ trace of $J$ and $N$ is the
smallest integer such that the next trace $J^{[N+1]}$ can not be
defined. The coefficients of this expansion depend on the spin $s$
of $J$ and on the space-time dimension $D$. They are determined by
the results collected in the Appendix, that lead to the one-term
recursion relation
\be \rho_{n+1}(D,s)\ =\ -\  {\rho_n(D,s) \over D+2(s-n-2)} \ ,
\label{rhon} \ee
with the initial condition $\rho_0(D,s)=1$. Notice that
$\rho_n(D-2,s)=\rho_n(D,s-1)$.

As we have seen, a generic tensor can be rendered transverse by the
application of the projector $\Pi$. In order to build the tensor
${\cal P}$ that defines the current exchange, one can first project
onto the transverse part of $J$ to then extract the traceless part
of the resulting tensor, thus obtaining
\be {\cal P}\; J \ = \ \sum_{n=0}^{N}\, \rho_n(D-2,s)\, \Pi^{\; n}\, J^{[n]}\ \label{loc1} . \ee
This is just what we have seen in eq.~(\ref{traceless}), but for a
key novelty: here $\rho$ depends on $D-2$, as a result of the
presence of $\Pi$.

We have thus made our way backwards to the tensor ${\cal P}$ of eq.~(\ref{compare}), that projects a covariant current $J$ onto its transverse
traceless part, but at the price of an explicit dependence on both the physical momentum $p$ and the additional vector $\bar{p}$. In some
notable cases, however, the dependence on $\bar{p}$ disappears. For instance, if $J$ is conserved,
\be J \cdot {\cal P} \cdot J \ = \ \sum_{n=0}^{N}\, \rho_n(D-2,s) \,
{s!\over {n!\,(s-2n)!\, 2^n}}\ J^{[n]} \cdot J^{[n]}, \label{loc11}
\ee
so that ${\cal P}$ can be effectively replaced by
\be {\cal P}_cJ\ =\ \sum_{n=0}^{N}\, \rho_n(D-2,s)\, \eta^n\,
J^{[n]}\ , \label{Ac1} \ee
and one can now show that
\be
\begin{split}
& ({\cal P}_cJ)^{\, \prime}\, = \, 2\, \sum_{0}^{N-1}\, \rho_{n+1}(D-2,s)\,
\eta^{\, n}\, J^{\, [n+1]}\, , \\
& ({\cal P}_cJ)^{\, \prime\prime}\ =\ 0\, .  \label{Ac2}
\end{split}
\ee

The momentum dependence of ${\cal P}$ can be also suppressed in
Fronsdal's constrained formulation \cite{fronsdal}. The Lagrangian
equation is in this case
\be \cF \ - \ \frac{1}{2} \ \eta \ \cF^{\; \prime} \ = \ J \
,\label{freq} \ee
and the double-trace condition may be used to show that
\be J^{\; \prime}\ =\ - \ \frac{1}{2\, \rho_1(D-2,s)} \ \cF^{\;
\prime}  \ \label{jfprime} , \ee
while the double trace of $J$ vanishes,
\be J^{\; \prime\prime} \ = \ 0 \ , \ee
so that eq.~(\ref{freq}) can be also presented in the alternative
form
\be \cF \ = \ J \ +\ \rho_1(D-2,s) \, \eta \, J^{\; \prime} \ .\ee
In addition, eq.~(\ref{freq}) implies that
\be p \cdot J \ = \ - \ \frac{1}{2} \ \eta \ p \cdot \cF^{\; \prime}
\ , \ee
so that, using eq.~(\ref{jfprime}), one can see that \emph{only} the
traceless part of the divergence of $J$ vanishes in general. The
last condition can be also written
\be p \cdot J\ +\ \rho_1(D,s-1)\ \eta \ p \cdot J^{\; \prime} \ =\ 0
\ , \ee
so that, introducing a new tensor $\tau$, the combination
\be J \ + \rho_1(D-2,s)\, \eta \, J'\ + \ p \, \tau \ee
is actually both \emph{traceless} and \emph{transverse} provided
$\tau$ is \emph{traceless} and satisfies the condition
\be p \cdot \tau \ =\ -\ \rho_1(D-2,s)\ J^{\; \prime} \ . \ee

However, $p \, \tau$ yields a vanishing result upon contraction with
$J$ since, as we have seen, the traceless part of $p \cdot J$ vanishes.
The end conclusion is that in Fronsdal's constrained theory ${\cal
P}J$ can be actually replaced with
\be {\cal P}_c\ J\ =\ J\ +\ \rho_1(D,s-1)\ \eta \ J^{\; \prime} \ ,
\ee
and therefore the corresponding current exchange is determined by
the relatively simple expression
\be J \cdot J \ +\ {\rho_1(D-2,s) \, s\, (s-1)\over 2}\ J^{\;\prime}
\cdot J^{\;\prime}\ . \ee

In order to compare with the unconstrained formulation of the
previous Section, let us begin by noticing that, once the third of
eqs.~(\ref{boseeomk}), the constraint $\cC=0$, is enforced, the
coupling to an external source is described by
\be \cA \ - \ \frac{1}{2} \ \eta \ \cA^{\; \prime} \ + \eta^2 \ \cB
= \ J \, \label{eqaJ} \ee
while the double trace ${\cal A}^{\; \prime\prime}$ vanishes
identically. As a result, the quantity
\be K=J\, -\, \eta^2\, \cB \ee
is somehow the counterpart of Fronsdal's current in this case, since
on shell $K^{\; \prime\prime}=0$ as a result of the condition
$\cA^{\; \prime\prime}=0$, that as we have stated follows from the
constraint $\cC=0$. This condition also determines $\cB$, and thus
the Lagrange multiplier $\beta$, in terms of $J$.

One can now write
\be K\ =\ J\ +\ \sum_{n=2}^{N}\, \sigma_n\, \eta^{n}\, J^{[n]}, \ee
with the coefficients determined recursively by the condition that
the double trace $K^{\; \prime\prime}$ vanish. This apparently
results in a three-term recursion relation,
\be \sigma_n\ +\ \bigl[D+2(s-n-3)\bigr]\, \biggl\{2\sigma_{n+1}\ +\
\bigl[D+2(s-n-4)\bigr]\, \sigma_{n+2}\biggr\}\ =\ 0 \ ,
\label{3term} \ee
with the conditions
\be \sigma_2\ \bigl[ D+2(s-3) \bigr] \, \bigl[ D+2(s-4) \bigr] \ =\
-\ 1,\ \ 2\, \sigma_2\ +\ \sigma_3\, \bigl[ D+2(s-5) \bigr] \ =\ 0\
\label{condsigma} . \ee
However, introducing
\be u_n\ = \ \sigma_n \ +\ \bigl[D+2(s-n-3)\bigr]\, \sigma_{n+1}\ ,
\ee
one can turn eq.~(\ref{3term}) into the simpler two-term recursion
relation,
\be u_{n+1}\ =\ -\ u_n\ {1\over \bigl[D+2(s-n-3)\bigr]}\ ,\label{re}
\ee
which, taking into account the value of $u_2$ determined by
eqs.~(\ref{condsigma}),
\be u_2 \ = \ \frac{1}{\bigl[ D+2(s-3) \bigr] \, \bigl[ D+2(s-4)
\bigr]} \ , \ee
is actually solved by
\be u_n=\rho_n(D-2,s)\ , \ee
where the $\rho_n$ are defined in eq.~(\ref{rhon}).

Making use of (\ref{re}), the defining relation for the $u_n$
becomes
\be u_n\ =\ \sigma_n\ -\ {u_n\over u_{n+1}}\, \sigma_{n+1}\ , \ee
or, in terms of $v_n=\sigma_n/u_n$,
\be v_{n+1}\ -\ v_n+1\ =\ 0 \ , \ee
whose solution is $v_n=v_2-(n-2)=-n+1$. In conclusion,
\be \sigma_n\ =\ (-n+1)\ \rho_n(D-2,s)\ . \ee

Since ${\cal A}$ is doubly traceless, eq.~(\ref{eqaJ}) can be turned
into
\be {\cal A} \ = \ K\ +\ \rho_1(D-2,s)\, \eta \, K^{\; \prime} \ ,
\ee
and one can verify that
\be K\ +\ \rho_1(D-2,s)\, \eta \, K^{\; \prime} \ = \ \sum_n\
\rho_n(D-2,s)\ \eta^{n}\ J^{[n]}\ . \ee
This determines ${\cal P}\; J$, and the current exchange is finally
\be \sum_{n=0}^{N}\ \rho_n(D-2,s) \ {s!\over {n!\; (s-2n)!\; 2^n}}\ J^{[n]}\cdot J^{[n]} \, \label{cexch}\ee
which agrees with eq.~(\ref{loc11}), with the correct number of
degrees of freedom, since the exchange involves, as expected, a pair
of traceless conserved currents built from the original conserved
current $J$.


\scss{Fermi fields in flat space}


In the previous Subsection we have recalled how, in $D$ dimensions,
the physical degrees of freedom carried by a massless symmetric
rank-$s$ tensor $\varphi_{\mu_1 \ldots \mu_s}$ fill a symmetric
\emph{traceless} rank-$s$ tensor in $D-2$ transverse dimensions. For
massless Fermi fields the situation is similar, if technically more
involved, so that a $D$-dimensional spinor-tensor field $\psi_{\mu_1
\ldots \mu_m}$ carries physical degrees of freedom corresponding to
an \emph{on-shell} $\gamma$-\emph{traceless} spinor-tensor in $D-2$
transverse dimensions. The on-shell condition is of course stronger
for a Fermi field: not only does it select a light-like momentum,
but via the Dirac equation it further halves the number of its
degrees of freedom. This is reflected in the familiar presence, in
spinor propagators, of the matrix $\psl$, that for a light-like
momentum $p$ has precisely this effect.

In discussing current exchanges for Fermi fields, it is useful to
begin by defining the projection of $\psi_{\mu_1 \ldots \mu_m}$ to
its $\gamma$-traceless part. One can verify that this is effected by
\be {\cal T}_m\psi\ =\ \psi\ +\ \sum_{n=1}^{N}\, \rho_{n}(D,m+1)\,
\bigl(\eta^n\psi^{[n]}\ +\ \eta^{n-1}\gamma \ \psisl^{[n-1]}\bigr)\,
, \label{gtrproj} \ee
where the coefficients $\rho_n(D,s)$ were introduced in the previous
Subsection, since they also determine the traceless projection for
bosonic fields.

Using ${\cal T}_m$, one can also construct the projectors to the
doubly and triply $\gamma$-traceless parts of $\psi$. The first is
simply the traceless projector that was introduced in the previous
Subsection, but here we can also relate it to the projector of
eq.~(\ref{gtrproj}), according to
\be {\cal T}_m^{(2)}\psi \ = \ \sum_n\rho_n(D,m)\, \eta^n\,
\psi^{[n]}\ =\ {\cal T}_m\, \psi\ +\ {1\over D+2m-2}\ \gamma\ {\cal
T}_{m-1}\, \psisl \ . \ee
In a similar fashion, one can define a triply $\gamma$-traceless
projector,
\bea
 {\cal T}_m^{(3)}\psi &=& {\cal
T}_m^{(2)}\, \psi\ +\ {1 \over D+2m-4}\ \eta \ {\cal
T}_{m-2}\ \psi'\nonumber\\
&=& \psi\ -\ \eta\, \gamma\, \Big({1\over
(D+2m-6)(D+2m-4)}\psisl'+\dots\Big)\ . \label{projf} \eea
In deriving these results, use has been made of the relations
\be \rho_n(D-2,m+1)\ =\ \rho_n(D,m)\ ,\quad \rho_n(D,m-1)\ =\
{\rho_{n+1}(D,m)\over \rho_1(D,m)}\ , \label{rhoprops} \ee
which follow from eq.~(\ref{rhon}), and of the initial condition
$\rho_0(D,s)=1$.

If the propagator is now denoted by $\frac{\psll}{p^2} \, {\cal B}$,
the previous considerations lead to
 \be {\cal B}\, {\cal J}\ =\ {\cal J}\ +\
\sum_{n=1}^{N}\rho_n(D-2,m+1) \Big(\Pi^n({{\cal J}})^{[n]} \ +\
\Pi^{n-1}\, \gamma\, \gamma \cdot ({\cal J})^{[n-1]}\Big)\ ,
\label{genfermi} \ee
which projects the external current ${\cal J}$ to its transverse and
$\gamma$-traceless part. This is achieved in two steps:  the current
${\cal J}$ is first projected to its transverse part ${\cal J}^T$
using the projector $\Pi$ introduced for bosonic fields, and then
the $\gamma$ trace is eliminated from the resulting expression via
eq.~(\ref{gtrproj}). Notice that, as in the bosonic case, the
presence of $\Pi$ brings about the replacement of $D$ with $D-2$ in
this expression.

Simplifications are again possible in special circumstances, and in
particular if ${\cal J}$ is transverse: eq.~(\ref{genfermi}) reduces
to
\be {\cal B}_c \, {\cal J}\ = \ {\cal J}\ +\
\sum_{n=1}^{N}\rho_n(D-2,m+1)(\eta^n{\cal J}^{[n]} \ +\
\eta^{n-1}\gamma\, \gamma \cdot {\cal J}^{[n-1]})\ , \ee
without the need for any explicit $\Pi$ projectors, and hence with
no explicit dependence on ${\bar p}$. Notice that ${\cal B}_c \,
{\cal J}$ can also be written in the form
\be {\cal B}_c \, {\cal J}\ =\ {\cal T}_m^{(2)}{\cal J}\ +\
\rho_1(D,m)\ \gamma \ \gamma \cdot {\cal T}_m^{(3)}{\cal J} \ ,
\label{fpr2} \ee
which makes it manifest that the current thus projected is
\emph{triply} $\gamma$-\emph{traceless}. This is the counterpart of
the condition (\ref{Ac2}) obtained for bosonic fields. As we shall
see shortly, this presentation of the result is particularly useful
when comparing with the propagator for the local unconstrained
formulation of the previous Section.

In the Fang-Fronsdal theory with an external source ${\cal J}$ the
field equation is
\be {\cal S} \ - \ \frac{1}{2} \ \eta \ {\cal S}^{\; \prime} \ - \
\frac{1}{2} \ \gamma \ssl \ = \ {\cal J} , \label{ffj} \ee
which via the Bianchi identity implies the condition
\be
\partial \cdot {\cal J} \ = \ - \ \frac{1}{2} \ \eta \ \partial \cdot
{\cal S}^{\; \prime} \ - \ \frac{1}{2} \ \gamma \ \partial \cdot
\ssl \ . \ee
As a result only the divergence of the $\gamma$-traceless part of
${\cal J}$ vanishes in general,
\be {\cal T}_{m-1}(\partial \cdot {\cal J})=0 \ . \ee
In addition, the Fang-Fronsdal constraint on the triple
$\gamma$-trace of $\psi$ implies that $\cS$, and hence ${\cal J}$ on
account of eq.~(\ref{ffj}), are also triply $\gamma$-traceless. In
analogy with what was done for bosons, eq.~(\ref{ffj}) can thus be
inverted and ${\cal B}\, {\cal J}$ can be replaced with
\be {\cal B}_c\, {\cal J}\ =\ {\cal J}\ +\ \rho_1(D-2,m+1)\, \Big(
\eta\, {\cal J}'\ +\ \gamma\ \gamma \cdot {\cal J}\Big)\ ,
\label{ffexch} \ee
which in four space-time dimensions agrees with the expression given
in \cite{fronsdal}.

In the unconstrained local formulation of the previous Section the
situation is similar to some extent, since once the third of
eqs.~(\ref{fermieomk}) is enforced, the first, the field equation
for $\psi$, takes the form
 \be
 {\cal W}\ -\ {1\over 2}\ \gamma\ \cWsl \ -\ {1\over 2}\ \eta\ {\cal W}^{\, \prime}\ =\ {\cal
 J}\ -\ {1\over 4}\ \eta\, \gamma \, {\cal Z}\ \equiv \ {\cal K}\ .
 \label{covfermi} \ee
Just as for bosons the gauge invariant tensor ${\cal A}$ was doubly
traceless on shell, so one can show that the gauge invariant
spinor-tensor $\cW$ is triply $\gamma$-traceless on shell,
\be \cWsl^{\, \prime} \ = \ 0 \ . \label{triplyh}
\ee
Hence, the same property holds for ${\cal K}$, so that the
combination of ${\cal Z}$ and ${\cal J}$ results in
\be
 {\cal K}\ =\ {\cal T}^{(3)}_{m}{\cal J}\ ,
\ee
that is therefore an effective Fang-Fronsdal current. One can then
solve eq.~(\ref{covfermi}), obtaining
\be
 {\cal W}\ =\ {\cal K}\ + \ \rho_1(D-2,m+1)\ [\gamma\ {\not{\! \!{\cal K}}}\ +\ \eta\ {\cal K}^{\, \prime} ] \ ,
\ee
and therefore the current-current amplitude is determined by
\be {\cal K}\ + \ \rho_1(D-2,m+1) \ [\gamma\ {\not{\! \!{\cal K}}}\
+\ \eta\ {\cal K}^{\, \prime} ] \ . \ee
In terms of ${\cal J}$, using (\ref{projf}), one can now conclude
that
\be {\cal W}\ =\ {\cal T}_m^{(2)}{\cal J}\ +\ \rho_1(D,m)\ \gamma\
\gamma \cdot{\cal T}_m^{(3)}{\cal J} \ . \ee
This expression is the direct counterpart of eq.~(\ref{ffexch}),
since, as we have stressed, $\rho_1(D,m)=\rho_1(D-2,m+1)$. It
finally shows that in the unconstrained formulation the current
exchange has the form (\ref{fpr2}), which indeed guarantees that the
correct number of degrees of freedom propagates on-shell.


\scss{Current exchanges in an $AdS$ background}


In the last two Subsections we have shown that the constrained and unconstrained formulations for Bose or Fermi fields in flat space time are
equivalent even in the presence of external currents. This result reflects the occurrence, in both settings, of tensors (${\cal F}$ and ${\cal
A}$, or ${\cal S}$ and ${\cal W}$) that on-shell are effectively subject to the (Fang-)Fronsdal ($\gamma$-)trace constraints and satisfy the
same Bianchi identities. We have also identified effective currents ($K$ or ${\cal K}$) that behave exactly like the Fang-Fronsdal currents. We
now want to show briefly how this structure carries over to $AdS$ backgrounds.

Let us begin by deriving the current exchange for unconstrained Bose fields in $AdS$. The key observation is that, when $\beta$ is on shell, the
first of eqs.~(\ref{adsequations}) takes the form
\be
 {\cal A}_L \, -\, {1\over 2}\, g \, {\cal A}_L^{\, \pe}\, +\, g^{\, 2} \, \cC_L \, = \, J \, ,
 \ee
 where $g$ denotes the $AdS$ metric and, again, ${\cal A}_L^{\, \prime\prime} = 0$.
 The same steps followed in the preceding Subsections for the
 flat-space analysis then lead to
 \be
 {\cal A}_L \, = \, \sum_{n=0}^{N} \r_n\, (D - 2,\, s) \, g^{\, n}\, J^{\, [n]}\ .
 \ee

In order to proceed further, it is very convenient to introduce the Lichnerowicz operator \cite{lich}, that in an $AdS$ background takes the
form
\be \Box_L \, \varphi \, = \, \Box \, \varphi\ + {1\over {L^2}}\,
\left[s \, (D + s - 2)\, \varphi \, - \, 2 \, g \, \varphi^{\, \pe} \right]\ , \ee
and is particularly convenient, since it commutes with contraction and covariant differentiation and allows one to write the $AdS$ Fronsdal
operator as
\be \Big[ \Box_L \, - \, {2\over L^2}\ (s - 1) \, (D + s - 3)\Big] \, \varphi \, - \,
\nabla \, \nabla \cdot \varphi\, +\,\nabla^{\, 2}\,
\varphi^{\, \prime}\ . \ee
When contracted with a conserved current, clearly only the first
term is relevant. As a result, the current exchange is finally
determined by
\be \sum_{n=0}^{N} \r_n\, (D - 2,\, s) \, {s! \over {n! \, (s - 2 n)!\,  2^{\, n}}} \
J^{\, [n]} \cdot \Big[ \Box_L \, - \, {2\over
L^2}(s - 1)(D + s - 3)\Big]^{-1}J^{\, [n]}\ , \ee
which reduces to the flat space amplitude in the limit $L \to
\infty$.

Fermi fields in $AdS$, as we have seen, entail a few additional complications, since the current is now subject to a modified conservation law,
\be \nabla \cdot {\cal J}\ +\ {1\over 2L}\ \gamma \cdot {\cal J}\ =\
0\ . \ee
When the auxiliary fields are on shell, the relevant field equation
reads
\be
 {\cal W}_L\ -\ {1\over 2}\ \gamma\ \cWsl \ -\ {1\over 2}\ \eta\
 {\cal W}_L^{\, \pe} \ =\ {\cal
 J}\ -\ {1\over 4}\ \eta\ \gamma\ {\cal Z}_L\ \equiv \ {\cal K}_L,
 \ee
with, again, the triple $\gamma$-trace constraint $\cWsl^{\, \prime}_L \ =\ 0$. As in the flat case, one can first solve for ${\cal Z}_L$ and
then obtain ${\cal W}_L$
 from
 \be
{\cal W}_L\ =\ {\cal K}_L\ -\ {1\over D+2(m-2)}\ \left[\gamma\,
{\not {\! \! {\cal K}}}_L\ +\ \eta\, {\cal K}^{\, \pe}_L\right]\ ,
 \ee
where the effective current ${\cal K}_L$ is subject to the triple $\gamma$-trace constraint ${\cal K}_L={\cal T}_m^{(3)}{\cal J}$, with ${\cal
T}_m$ defined as in (\ref{projf}) but for the replacement of the flat metric and the Dirac matrices with their $AdS$ counterparts. Notice that
${\cal K}_L$ satisfies the modified conservation law
\be {\cal T}_m\, \Big(\nabla \cdot {\cal K}_L \ +\ {1\over 2L}\
{\not {\! \! {\cal K}}}_L\Big)\ =\ 0\ .
 \ee

One thus faces, again, an effective current ${\cal K}_L$ which is
built from the conserved current ${\cal J}$ but nonetheless behaves
as a constrained current of the Fang-Fronsdal theory. Constrained
and unconstrained local formulations agree, the latter being
effectively a Fang-Fronsdal theory with a partially conserved
current which is built, as stressed above, from the conserved
current ${\cal J}$.


\scs{Current exchanges in the non-local formulation}


We can now turn to the non-local formulation. Confining our
attention to bosonic fields, we begin by analyzing the current
exchange in the non-local theory with reference to the Lagrangian
equations proposed in \cite{fs1}, in order to display the problem.
As anticipated, and as we shall see shortly, the Lagrangian equation
proposed in \cite{fs1} \emph{is not} the proper counterpart of the
local one discussed in Section 2 since, with standard couplings of
the $\varphi \cdot J$ type, it does not result in the correct
counting of degrees of freedom in current exchanges. We shall then
present a unique set of non-local Lagrangian equations that, for all
$s$, reproduce the current exchange of the previous Section. The
result of this analysis will thus be a precise map, for all $s$,
between the local and non-local formulations of the unconstrained
theory. In the absence of external currents, however, these novel
Lagrangian equations can also be turned into the non-Lagrangian
equations of \cite{fs1}, eqs.~(\ref{oddcurv}) and (\ref{evencurv}).
Here we confine our attention, for brevity, to the case of bosonic
fields, but similar results are expected to hold for fermionic
fields.


\scss{The problem}


The iterative procedure of \cite{fs1}, reviewed in the previous
Section, was meant to terminate, for a given value $s$ of the spin,
$s=2n-1$ or $s=2n$, after first reaching a non-local gauge invariant
extension ${\cal F}^{(n)}$ of the Fronsdal operator. From this, as
shown in \cite{fs1}, one can build a divergence-free Einstein-like
tensor by simply combining ${\cal F}^{(n)}$ with its traces
according to
\be {\cal G}^{(n)} \ = \ \sum_{p=0}^n \ \frac{(-1)^p\, (n-p)\;
!}{2^p \ n\; !} \ \eta^{\, p} \ {\cal F}^{\, (n)\, [p]} \ . \ee
The form of this Einstein-like tensor is determined by the Bianchi
identity satisfied by the $\cF^{(n)}$,
\be \prd {\cal F}^{(n)} \ - \ \frac{1}{2n} \ \pr {{\cal F}^{(n)}}{\; '} \ = \ - \ \left( 1 + \frac{1}{2n}  \right) \ \frac{\pr^{\;
2n+1}}{\Box^{\; n-1}} \ \vf^{\, [n+1]} \ , \label{bianchin} \ee
where for the two relevant values of $s$ associated to a given $n$, the term on the right-hand side vanishes identically. In this case
eq.~(\ref{bianchin}) has a number of interesting consequences, that can be derived taking successive traces:
\be \prd {\cal F}^{(n)\, [k]} \ - \ \frac{1}{2(n-k)} \ \pr {{\cal
F}^{(n)\, [k+1]}} \ = \ 0 \ , \qquad ( k \leq n-1)\ .
\label{bianchink} \ee
In particular, if the spin $s$ is odd, so that $s=2n-1$, one can see
that
\be \prd {\cal F}^{(n)\, [n-1]} \ = \ 0 \ . \label{lastodd} \ee

If the system is coupled to an external current ${\cal J}$, the
Lagrangian field equations proposed in \cite{fs1} thus read
\be {\cal G}^{\, (n)} \, \equiv \, \sum_{p=0}^{n}{(-1)^p \,
\frac{(n-p)!}{2^p \, n!}}\ \eta^{\, p} \ {\cal F}^{\, (n)\, [p]}\,
=\, {\cal J}\ , \label{nloc1} \ee
where, as we have anticipated, $s=2n-1$ or $s=2n$. These equations
can now be inverted noticing that
\be \rho_1(D-2n,s)\, \eta \, {\cal J}^{\, \prime} \, =\, - \,
\sum_{p=1}^{n}{(-1)^p \ \frac{p\; (n-p)!}{2^p n!}}\ \eta^{\, p}\,
{\cal F}^{\, (n)\, [p]}\ , \ee
\be \rho_2(D-2 n,s) \, \eta^2{\cal J}^{\; \prime\prime}\ =\
\sum_{p=2}^{n}(-1)^p \, {p(p-1)\over 2} \, {(n-p)!\over {2^p n!}}\
\eta^{\; p}\ {\cal F}^{(n)[p]}\ , \ee
and continuing in this fashion one readily obtains
\be {\cal F}^{(n)}\ =\ \sum_{n=0}^{n}\ \rho_n(D-2n,s)\ \eta^n\ {\cal
J}^{[n]}, \label{nloc2} \ee
where the coefficients $\rho_n$ were introduced in Section 3.

One should now contract eq.~(\ref{nloc2}) with a conserved current,
and the occurrence of a key simplification may be simply
anticipated: the relations ${\cal J} \cdot {\cal F}^{(k)}=\ldots
={\cal J} \cdot {\cal F}$  hold, up to terms involving the vanishing
divergence of ${\cal J}$. The current exchange in the non-local
theory of \cite{fs1}, based on the Einstein-like tensor of
eq.~(\ref{nloc1}), is thus determined by
\be \sum_{n=0}^{n}\rho_n(D-2n,s)\, \eta^n\, {\cal J}^{[n]} \ ,
\label{nloc3} \ee
and clearly \emph{disagrees} with eq.~(\ref{loc1}) whenever $s>2$,
\emph{i.e.} for all interesting cases of higher-spin fields.

In order to better appreciate the nature of the problem, it is
instructive to take a closer look at the relatively simple but still
non-trivial case of a spin-3 field. The kinetic operator of
\cite{fs1},
\be {\cal F}^{\, (2)}_{\mu\, \nu\, \rho} \ = \ {\cal F}_{\mu\, \nu\, \rho} \, - \, \frac{1}{3\, \Box} \, \left( \partial_{\, \mu} \,
\partial_{\, \nu}\,  {\cal F}^{\, \prime}_{\, \rho} \, + \, \partial_{\, \nu} \, \partial_{\, \rho}\,  {\cal F}^{\, \prime}_{\, \mu} \, +\,
\partial_{\, \rho}\,  \partial_{\, \mu}\,  {\cal F}^{\, \prime}_{\, \nu} \right)\, ,\ee
is then determined by a single iteration of eq.~(\ref{recursion}),
and satisfies the Bianchi identity
\be
\partial \cdot {\cal F}^{\, (2)} \ - \ \frac{1}{4}\ \partial \, {\cal F}^{\, (2)\,
\prime} \, = \, 0 \ , \label{bianf1} \ee
to be compared with the conventional Bianchi identity for the
Fronsdal operator
\be
\partial \cdot {\cal F} \, - \, \frac{1}{2} \, \partial \, {\cal F}^{
\, \prime} \, = \, 0 \, . \label{bianf} \ee
In both cases the Lagrangian field equations would couple the
corresponding divergence-free Einstein-like tensors to
divergence-free currents, according to
\bea &&{\cal G} \ \equiv \ {\cal F} \ - \ \frac{1}{2} \ \eta \ {\cal
F}^{\; \prime} \ = \ J \ , \label{gj} \\
&&{\cal G}^{(2)} \ \equiv \ {\cal F}^{\, (2)} \ - \ \frac{1}{4} \ \eta \ {\cal F}^{(2)\; \prime} \ = \ {\cal J} \ , \label{g1j1} \eea
but the definitions of ${\cal G}$ and ${\cal G}^{(1)}$ involve
different coefficients, reflecting the differences between the
Bianchi identities of eqs.~(\ref{bianf1}) and (\ref{bianf}).
Inverting these equations, one would then arrive, in the two cases,
at the current exchanges
\bea
&& J \cdot J \ - \ \frac{3}{D} \ J^{\; \prime} \cdot J^{\,
\prime}\ , \label{locex} \\
&& {\cal J} \cdot {\cal J} \ - \ \frac{3}{D-2} \ {\cal J}^{\;
\prime} \cdot {\cal J}^{\, \prime} \ , \label{nlocex}
\eea
which are clearly different, as special cases of eqs.~(\ref{loc1})
and (\ref{nloc3}). The lesson to be drawn from this example is that
the differences between the current exchanges in the local
formulation of the previous Sections and in the non-local
formulation of \cite{fs1} based on eq.~(\ref{nloc1}) reflect those
between the modified Bianchi identity satisfied by the non-local
kinetic operators ${\cal F}^{(n)}$ and the original Bianchi identity
satisfied by the Fronsdal operator ${\cal F}$.

Interestingly, the two results obtained in this spin-3 example can
actually be mapped into one another provided the currents $J$ and
${\cal J}$ are related by a suitable \emph{non-local} field
redefinition. Indeed, defining
\be \bar{\eta} \ = \ \eta \ - \ \frac{\partial^2}{\Box} \ , \ee
a non-local extension of $\eta$ which is \emph{divergence free}, if ${\cal J}$ and $J$ are related according to
\be {\cal J} \ = \ J \ + \ \frac{-3 D \pm \sqrt{3 D^2 - 6 D}}{3 D
(D+1)} \ \bar{\eta} \ J^{\; \prime} \ ,\ee
eq.~(\ref{nlocex}) turns precisely into (\ref{locex}). Notice that
the map between the two constructions thus obtained is compatible
with the conservation of both $J$ and ${\cal J}$.

This result is adding a useful piece of information: while the
tensor ${\cal G}^{(1)}$ does not couple as expected to an external
current, a suitable non-local combination of this tensor with its
trace does. These more singular objects satisfy Bianchi identities
that are closer to that satisfied by ${\cal F}$ and ${\cal A}$:
hence, they are precisely the types of non-local kinetic operators
we are after, for all values of $s$. As a side technical remark, let
us stress again that, strictly speaking, neither $\bar{\eta}$ nor
the ${\cal F}^{(n)}$ are well-defined quantities on shell. Hence, it
should be understood that, in all the present treatment of current
exchanges in the non-local formulation, one is actually working off
shell all the way and only the final results are continued on-shell
to arrive at the correct counting of degrees of freedom.

Actually, in \cite{fs1} a more singular class of non-local operators
was also considered. For all $s$, they can be obtained combining the
${\cal F}^{(n)}$ with their traces and are quite interesting, since
they lead in general to field equations of the type
\be \tilde{\cal F} \, \equiv \, {\cal F} \, - \, 3 \, \partial^{\,
3} \, {\cal H} \ = \ 0 \ , \ee
that have the same form as the non-Lagrangian compensator equations
(\ref{nonlagr}). In particular, for $s=3$ one can define
\be \tilde{{\cal F}}^{(1)}_{\mu\nu\rho} \ \equiv \ {\cal
F}_{\mu\nu\rho} \ - \ \frac{\partial_\mu \partial_\nu
\partial_\rho}{\Box^2} \
\partial \cdot {\cal F}^{\; \prime} \ , \label{doublepole}
 \ee
which actually satisfies the same Bianchi identity as ${\cal F}$, so
that the corresponding Einstein tensor is
\be \tilde{\cal G} \ = \ \tilde{\cal F} \ - \ \frac{1}{2} \ \eta \
\tilde{\cal F}^{\, \pe} \ . \label{bianf2} \ee

Since ${\cal F}$ and $\tilde{\cal F}$ differ by terms that vanish
upon contraction with a conserved current, with this type of
non-local Lagrangian equation the current exchange clearly agrees
with the result obtained in the previous Subsection for the local
theory. Notice that in this case one could also arrive at
eq.~(\ref{doublepole}) starting from the two conditions
\bea && {\cal A} \ - \ \frac{1}{2} \ \eta \ {\cal A}^{\; \prime} \ =
\ 0  \nonumber \\
&& \partial \cdot {\cal A}^{\; \prime} \ = \ 0 \ , \eea
where the second guarantees that the first couple consistently to a
conserved current.

The lesson to be drawn from this example is, therefore, that for a
given $s$, equal to $2n-1$ o $2n$, the Einstein-like tensors that
correspond to those of the local theory \emph{are not} directly
expressible in terms of the single ${\cal F}^{(n)}$ operator, but
are to involve other more singular operators ${\cal F}^{(m)}$, with
$m > n$. The issue is now to determine the form of these
Einstein-like tensors for all $s$ and to elucidate the corresponding
geometrical structure. As in other circumstances, for instance for
the Lagrangians of Section 2, one has to reach the $s=4$ case to
first uncover the relevant pattern. In this case, letting
\be {\cal A} \, = \, {\cal F} \, - \, 3\, \partial^{\, 3} \,
\alpha_{\, \varphi} \, ,  \nonumber \ee
and insisting on the condition $\partial \cdot {\cal A}^{\; \prime}
= 0$, leads to
\be \a_{\, \vf} \, = \, \fr{1}{3 \, \Box^{\, 2}} \, \prd \cF^{\,
\pe} \, - \, \fr{\pr}{4 \, \Box^{\, 3}} \, \prd \prd \cF^{\, \pe} \,
. \ee
On the other hand, combining it with its trace, one can turn the
non-local equation ${\cal F}^{(1)}=0$ into
\be {\cal T}\, \equiv \, {\cal F} \, - \, 3\, \partial^{\, 3}\,
{\cal H}_{\, \varphi} \, = \, 0 \, , \ee
where
\be \cH_{\, \vf} \, = \, \fr{1}{3 \, \Box^2} \, \prd \cF^{\, \pe} \,
- \, \fr{\pr}{4\, \Box^{\, 2}} \, \cF^{\, \pe \pe}\, , \ee
Notice that $\alpha_{\, \varphi}$ and ${\cal H}_{\, \varphi}$
\emph{do not coincide}. Their difference, however, is determined by
a gauge invariant quantity according to
\be \a_{\, \vf} \, - \, \cH_{\, \vf} \, = \, \fr{3}{4} \,
\fr{\pr}{\Box} \, (\vf^{\, \pe \pe} \, - \, 4 \, \prd \a_{\, \vf}) \
, \ee
a relation that can also be presented in the more symmetric form
\be \vf^{\, \pe \pe} \, - \, 4 \, \prd \a_{\, \vf} \ = \ \frac{1}{4}
\left[ \vf^{\, \pe \pe} \, - \, 4 \, \prd \cH_{\, \vf} \right] \
,\ee
and implies that the two choices result in the two inequivalent
Einstein tensors
\be
\begin{split}
E_{\, \a}   \, &= \, \cA \, - \, \12 \, \h \, \left( \cA^{\, \pe} \,
- \, \fr{1}{3} \,
 \fr{\pr^2}{\Box} \, \cA^{\, \pe \pe} \right) \, - \, \fr{1}{6} \, \h^2 \, \cA^{\, \pe \pe} \, ,  \\
E_{\, \cH} \, &= \, {\cal T} \, - \, \12 \, \h \, \left( {\cal
T}^{\, \pe} \, - \, \fr{1}{3} \,
 \fr{\pr^2}{\Box} \, {\cal T}^{\, \pe \pe} \right) \, + \, \fr{5}{24} \, \h^2 \, {\cal T}^{\, \pe \pe} \
 .
\end{split}
\ee
Notice that none of the two choices results in the correct current
exchange. The first indeed gives
\be {\cal J} \cdot \cA \ =\ {\cal J} \cdot {\cal J} \ - \
\frac{6}{D+2} \ {\cal J}^{\; \prime} \cdot {\cal J}^{\; \prime} \ +
\ \frac{3(D+6)}{(D+2)(D^2+6 D + 2)}\  \left({\cal J}^{\;
\prime\prime}\right)^2 \ , \ee
while the second gives
\be {\cal J} \cdot {\cal T} \ =\ {\cal J} \cdot {\cal J} \ - \
\frac{6}{D+2} \ {\cal J}^{\; \prime} \cdot {\cal J}^{\; \prime} \ +
\ \frac{3(6 - 5 D)}{(D+2)(-5 D^2+6 D + 8)}\  \left({\cal J}^{\;
\prime\prime}\right)^2 \ , \ee
to be compared with the correct result, given in eq.~(\ref{loc11}).

\scss{Nonlocal equations with a proper current exchange}

The arguments of the previous Subsection show clearly that the
Einstein tensor is the crucial ingredient behind the current
amplitude, whose form is determined by the Bianchi identity for
$\cA_{nl}$, the gauge-invariant extension of the Fronsdal operator
$\cF$. Actually, a closer look at the discussion of Section 3 for
the local theory shows that the non-local formulation can reproduce
the correct current-current amplitude provided its Lagrangian field
equations
\be \cG_{nl} \ =\ \cJ \label{nlocleq} \ee
lead to the solution
\be \cA_{nl}\ =\ {\cal P}_c\cJ \ , \label{nloclsol} \ee
where ${\cal P}_c$ was defined in eq.~(\ref{Ac1}). Eqs.~(\ref{Ac2}),
however, imply a pair of consistency conditions for this statement,
\bea
&& \cA_{nl}^{\; \prime\prime} \ = \ 0 \ , \nonumber \\
&& \cA_{nl}\ -\ {1\over 2}\ \eta\, \cA_{nl}^{\; \prime} \ =\ \cJ\ +\
\eta^2(\ldots) \ , \eea
which, as we have seen, are precisely met by the local construction.

One is thus led to search for an Einstein tensor that differs from
that of eq.~(\ref{nloc1}), and is rather of the form
\be \cG_{nl}\ =\ \cA_{nl}\ -\ {1\over 2}\, \eta\, \cA_{nl}^{\;
\prime} \ +\ \eta^2\,  \cB \ ,\label{enl} \ee
for some tensor $\cB$, and where $\cA_{nl}''=0$.
Conversely, given this form of the Einstein tensor, with a doubly
traceless $\cA_{nl}$, one can see that the solution of
eq.~(\ref{nlocleq}) is bound to take the form (\ref{nloclsol}). On
the other hand, the Einstein tensor can take the form of
eq.~(\ref{enl}) only if $\cA_{nl}$ obeys the Bianchi identity
\be \prd \cA_{nl}\ -\ {1\over 2}\ \partial\cA_{nl}^{\; \prime}\ =\
0\ ,\label{bbb} \ee
and if the divergence of the trace of $\cA_{nl}$ is a pure gradient,
a condition that we shall write in the form
\be \prd \cA^{\, \pe}_{nl} \ =\ 2\, \pr \, \cD_{nl}\ ,\label{b2} \ee
with ${\cal D}_{nl}$ an arbitrary non-local tensor. These two
requirements are in fact necessary to guarantee the consistency of
eq.~(\ref{nlocleq}) with the conservation of the current ${\cal J}$.

Eqs.~(\ref{bbb}) and (\ref{b2}) are also sufficient if they are
combined with the additional demand that $\cA_{nl}$ be of the form
\be \cA_{nl}\ =\ \cF\ -\ 3\,
\partial^{\, 3}\, \a_{nl}, \label{anlf} \ee
for some $\alpha_{nl}$ to be determined. In order to prove that this
is actually the case, notice first that the Bianchi identity reads
\be \pr \cdot \cA_{nl} \ - \ \frac{1}{2} \, \pr \, \cA^{\;
\prime}_{nl} \ = \ - \ \frac{3}{2} \ \pr^3 \, \left( \vf^{\;
\prime\prime} \, - \, 4 \pr \cdot \a_{nl} \, - \, \pr \, \a^{\;
\prime}_{nl} \right)\ , \label{bianchinl} \ee
so that it vanishes provided $\alpha_{nl}$ solves the constraint
\be \cC_{nl}\ \equiv \ \left( \vf^{\; \prime\prime} \ - \ 4 \pr
\cdot \a_{nl} \ - \ \pr \, \a^{\; \prime}_{nl} \right)\ =\ 0\ .
\label{bnl} \ee
Since the double trace of $\cA_{nl}$ can be expressed as
\be \cA^{\pe \pe}_{nl}\ =\ 3\, \Box \, \cC_{nl} \ +\ 3\,
\partial \, \prd \cC_{nl}\ +\
\partial^{\, 2} \, \cC_{nl}^{\, \prime} \ , \label{dtracenl}
\ee
the condition (\ref{bnl}) also guarantees that $A_{nl}$ be doubly
traceless on shell. However, eq.~(\ref{b2}) and its successive
traces also imply that
\be \cG_{nl}\ =\ \cA_{nl} \ - \ \frac{1}{2} \ \eta \,
\cA^{\, \pe}_{nl} \  +\ \eta^{\, 2} \,  {\cD_{nl}}\ +\ \ldots+
\eta^{\, n+2}\, {\cD_{nl}^{\, [n]}\over 2^{n}\, n!}, \ee
where $n$ is the integer such that $s=2(n+2)$ or $s=2n+5$, is indeed
conserved and is of the form (\ref{enl}).

To summarize, the current-current amplitude is correctly reproduced
by the modification (\ref{anlf}) of the Fronsdal operator provided
the two conditions (\ref{bnl}) and (\ref{b2}) are met. Notice that
(\ref{bnl}) and its consequence (\ref{dtracenl}) are the
counterparts of the $\beta$ equation of motion in the local
formulation (the third equation in (\ref{boseeomk})). On the other
hand, eq.~(\ref{b2}) can be regarded as the counterpart of the $\a$
equation of motion of the local unconstrained formulation, while the
very form of the Einstein tensor (\ref{enl}) is clearly tailored
after the one entering the $\vf$ equations of motion in the local
formulation.

Now we would like to show that a solution to these two conditions
exists and is unique. To this end, we shall first determine
$\a_{nl}$ in terms of $\vf$. We shall then conclude the present
Subsection by displaying the geometrical structures underlying
$\cA_{nl}$.

Let us first notice that
\be
\begin{split}
 \prd \cA^{\, \pe}_{nl} \, = & \, 3 \, \Box \, \prd \vf^{\, \prime} \ -
 \,  2\, \prd \prd \prd \vf\ -\ 3\, \Box^{\, 2} \, \a_{nl} \\
&-\, \pr\, \biggl(9\, \Box \, \prd \a_{nl}\ +\ 3\, \pr \, \prd \prd \a_{nl} \\
& + \, {3\over 2}\ \Box\, \pr\, \a^{\, \pe}_{nl}\ +\ \pr^{\, 2} \, \prd \a^{\, \pe}_{nl} \ - \
\prd \prd \vf^{\, \prime}\ -\, \Box\, \vf^{\, \prime\prime}\ -\
 {1\over 2}\ \prd\,
\vf^{\, \prime\prime} \biggr) \ .
\end{split}
\ee
Requiring that this expression be equal to $2\, \pr \, {\cD}_{nl}$
gives the two conditions
\bea
&& 3\, \Box\, \prd \vf^{\, \prime} \ -\ 2\prd \prd \prd \vf\ -\ 3\, \Box^{\, 2} \, \a_{nl}\
 =\ \pr \, f \nonumber \\
&&\pr \, \left(9\, \Box\,\prd \, \a_{nl}\ +\ 3\pr\, \prd \prd \a_{nl}\ + \
{3\over 2}\ \Box\, \pr \, \a^{\, \prime}_{nl}\ \right. \nonumber \\
&& +\ \left. \pr^{\, 2}\, \prd \a^{\, \prime}_{nl}\ -\ \prd \prd \vf^{\,
\prime} \ -\ \Box\, \vf^{\, \prime\prime}\ -\ {1\over 2}\ \pr\, \prd
\vf^{\, \prime\prime}\right)\ = \ f\ - \ 2\, \cD_{nl} \ ,
\label{dnl}\eea
where $f$ is an arbitrary tensor, that will be determined shortly by
eq.~(\ref{bnl}). The first equation gives indeed
\be \a_{nl}\ =\ {1\over \Box}\ \prd \vf^{\, \prime} \ -\ {2\over
3 \, \Box^{\, 2}} \, \prd \prd \prd \vf \ -\ {\pr\over 3\, \Box^{\, 2}}f\ , \ee
and inserting this expression in eq.~(\ref{bnl}) yields an equation
for $f$,
\be
\begin{split}
f &= \, -{3\over 4}\ \Box\, \vf^{\, \prime\prime }\ -\ {2\over
\Box}\ \prd \prd \prd \prd\vf\ +\ {3}\, \prd \prd \vf^{\, \prime} \\
&+ \, {3\over 4}\ \pr \, \prd \vf^{\, \prime\prime}\ - \ {1\over
2\Box}\ \pr \, \prd \prd \prd \vf^{\, \prime} \ -\ {3\over 2}\
{\pr\over\Box}\ \prd f\ -\ {1\over 2}\ {\pr^{\, 2} \over \Box}\ f^{\,
\prime}\ .
\end{split}
\ee
One can now look for a solution of the form
\be f=\sum \partial^{\; n}\, f_n \ , \ee
where for a spin-$s$ field the sum terminates at $n=s-4$, to be
determined by successive iterations, and the result is
\bea f_0&=&-\ {3\over 4}\ \Box\, \vf^{\; \prime\prime}\ -\
{2\over\Box}\ \prd \prd \prd \prd \vf\ +\
{3}\, \prd \prd \vf^{\; \prime} \\
f_1&=& {3\over 4}\ \prd \vf^{\; \prime\prime}\ -\ {2\over \Box}\
\prd \prd \prd \vf^{\; \prime}\ +\ {6\over 5 \, \Box^2}\
\prd \prd \prd \prd \prd \vf\\
f_n&=&-\ {1\over (n+1) \, (n+4)}\ {1\over\Box}\ \Big({2\, n \, (n+2)} \, \prd
f_{n-1} \ +\ {n \, (n-1)} \, f^{\, \pe}_{n-2}\Big)  \quad (n\ge 2) \ . \eea
This truncation has an interesting consequence: if $\a_s$ denotes
the expression for $\a_{nl}$ for a spin $s$ field, taking into
account the last terms leads to the recursion relation
\be \a_{s+1}\ =\ \a_{s}\ -\ {s-2 \over 3}\ {\pr^{\, s-2}\over \Box^2}\
f_{s-3} \ . \ee
The solutions for the first few cases are
\bea
\alpha_3&=&{1\over \Box}\ \prd \vf^{\, \prime}\ -\ {2\over 3 \, \Box^{\, 2}}\ \prd \prd \prd \vf\\
\alpha_4&=&\alpha_3\ +\ {\pr\over \Box^{\, 2}}\ \Big({1\over 4}\
\Box \, \vf^{\, \prime\prime}\ + \ {2\over3 \, \Box}\ \prd \prd \prd \prd
\vf\ +\
\prd \prd \vf^{\; \prime} \Big)\\
\alpha_5&=&\alpha_4\ -\ {\pr^{\, 2} \over\Box^{\, 2}}\ \Big({1\over 2}\ \prd
\vf^{\; \prime\prime}\ - \ {4\over 3 \, \Box}\ \prd \prd \prd \vf^{\,
\prime}\ +\ {4\over 5 \, \Box^{\, 2} } \ \prd \prd \prd \prd \prd \vf\Big) \ ,
\eea
and can also be expressed in terms of the Fronsdal operator as
\bea
\alpha_3&=&{1\over3 \, \Box^{\, 2}}\ \prd \cF^{\, \prime} \ ,\\
\alpha_4&=&{1\over3 \, \Box^{\, 2}}\ \prd \cF^{\, \prime} \ -\ {\pr\over
3 \, \Box^{\, 3}}\ \prd \prd \cF^{\, \prime}\ +\ {1\over 12}\ {\pr\over \Box^{ \, 2} }
\, \cF^{\, \prime\prime}\ ,\\
\alpha_5&=&{1\over3 \, \Box^{\, 2}} \, \prd \cF^{\, \prime}-{\pr\over
3\, \Box^{\, 3}}\pr\cdot\pr\cdot\cF^{\, \prime}+{2\over 5}{\pr^{\, 2} \over
\Box^{\, 4}}\prd\prd\prd\cF^{\; \prime} +{1\over 12}{\pr\over
\Box^{\, 2}} \, \cF^{\, \prime\prime}-{1\over 5}{\pr^{\, 2} \over
\Box^{\, 3}} \, \pr\cdot\cF^{\, \prime\prime}. \eea
The Einstein tensor is finally determined by $\a_{nl}$ and
$\cD_{nl}$ given in (\ref{dnl}). For spin $s=4$ and $s=5$, for
instance, the results read
\bea
\cD_4&=&{1\over 2}\left({1\over \Box}\pr\cdot\pr\cdot\cF^{\; \prime}\ -\ \cF^{\; \prime\prime}\right)\\
\cD_{5}&=&{1\over 2}\left({1\over \Box}\ \pr\cdot\pr\cdot\cF^{\;
\prime}\ -\ \cF^{\; \prime\prime}\ -\ {\pr\over \Box^2}\
\prd\prd\prd\cF^{\; \prime}\  +\ {\pr\over \Box}\ \prd \cF^{\;
\prime\prime}\right)\nonumber\\&=&{1\over2} \left(1\ -\
{\pr \, \prd\over \Box}\right) \left({1\over \Box}\
\pr\cdot\pr\cdot\cF^{\; \prime}\ -\ \cF^{\; \prime\prime}\right).
\eea

There is actually a more illuminating way to proceed. The key idea
is to present the solution in a manifestly gauge invariant fashion
by expressing $\cA_{nl}$ in terms of the higher-spin curvatures. To
this end, let us begin by considering the geometric operators
 \be \label{kinetic}
   \cF^{(n+1)} \, =
     \begin{cases}
    \fr{1}{\Box^{\, n}} \, \cR^{\, [n+1]} \, & \, s \, = \, 2\, (n \, + \, 1) \, , \\
    \fr{1}{\Box^{\, n}} \, \prd \cR^{\, [n]} \, & \, s \, = \, 2\, n \, + \, 1 \,
    .
    \end{cases}
 \ee
As discussed in \cite{fs1}, these operators are directly related to
the $\cF^{(n)}$ of eq.~(\ref{recursion}), and hence satisfy the
sequel of identities
\be \prd \cF^{(n+1)\; [k]} \, = \, \fr{1}{2\, (n \, - \, k \, + \, 1)} \, \pr \, \cF^{(n+1)\; [k + 1]} \ ,  \label{bianchinn} \ee
so that, in the odd case, $\prd \cF^{(n+1)\; [n]} \, \eq \, 0$.

As a result, all divergences of the tensors $\cF_{n+1}$ can be
expressed in terms of traces, so that the only available independent
structures are
\be \cF^{(n+1)} , \ \ \ \cF^{(n+1)\; \pe} , \ \ \ \cF^{(n+1)\; \pe \pe} , \ \ \ \dots ,\ \ \ \cF^{(n+1)\; [k]} , \ \ \ \dots ,\ \ \ \cF^{(n+1)\;
[q]} \, , \ee
where $q \, = \, n \, + \, 1$ or $q \, = \, n$ depending on whether
the rank is $s \, = \, 2\, (n \, + \, 1)$ or $s \, = \, 2\, n \, +
\, 1$. One can now construct the most general linear combination of
\emph{all} these terms, with arbitrary coefficients (the first may
be set to one, up to an overall normalization)
\be \label{geomtens} \cA_{nl} \, =  \, \cF^{(n+1)} \, + \, a_1 \, \fr{\pr^{\, 2}}{\Box} \, \cF^{(n+1)\; \pe} \, + \, \dots \, +\, a_k \,
\fr{\pr^{\, 2k}}{\Box^{\, k}} \, \cF^{(n+1)\; [k]} \, + \, \dots \, + \,
\begin{cases}
 a_{n+1} \, \fr{\pr^{\, 2\, (n+1)}}{\Box^{\, n+1}} \, \cF^{(n+1)\; [n+1]} & s \, = \, 2\, (n + 1) \, ,\\
 a_{n} \, \fr{\pr^{\, 2\, n}}{\Box^{\, n}} \, \cF^{(n+1)\; [n]} & s \, = 2\, n + 1 \, .
\end{cases}
\ee
One can first determine $a_1$ by requiring that $\cA_{nl}$ be of the
form (\ref{anlf}). The remaining coefficients may then be fixed,
recursively, demanding that the proper Bianchi identity,
\be \prd \cA_{nl}\ -\ {1\over 2}\ \pr\, \cA_{nl}^{\; \prime} \ =\ 0
\ , \label{bianchiok} \ee
hold for $A_{nl}$ or, equivalently, that $A_{nl}$ be doubly
traceless. Indeed, demanding that all terms in $\pr^{\, 2}$
disappear yields the condition
\be a_1 \, = \, \fr{n}{n \, + \, 1} \ , \ee
while the double trace $\cA^{\; \prime\prime}$ vanishes provided
\be a_{k+2} \, = \, - \, \fr{n \, + \, k + 1}{n \, - \, k}\, \left\{
 \fr{n \, + \, k}{n \, - \, k \, + \, 1}\, a_k \,+ 2 \, a_{k + 1} \right\}
 \ ,
\ee
so that there is one and only one solution to our conditions.

The explicit solution of this difference equation may be foreseen
after a few iterations, and reads
 \be a_k \, = \, (-1)^{k+1} \, (2\, k\,
-\, 1)\, \fr{n\, + \, 2}{n\, - \, 1} \, \prod_{j= -1}^{k-1}\, \fr{n
\, + \, j}{n \, - \, j \, + \, 1} \, , \label{solak} \ee
so that
\be \cA_{nl} \, = \, \sum_{k=0}^{n+1} \, (-1)^{k+1} \, (2\, k\,  -\, 1)\, \left\{ \fr{n\, + \, 2}{n\, - \, 1} \, \prod_{j=0}^{k-1}\, \fr{n \, +
\, j}{n \, - \, j \, + \, 1} \right\} \, \fr{\pr^{\, 2k}}{\Box^{\, k}}\, \cF^{(n+1) \; [k]} \ . \ee
Alternatively, the Bianchi identity (\ref{bianchiok}) holds provided
\be \label{bianchicoeff} a_{k+1} \ = \ - \ a_k \ \frac{(2 k + 1)(n +
k)}{(2 k -1)(n -k +1)} \ ,\ee
which is also solved by eq.~(\ref{solak}).

It is remarkable that (\ref{bianchicoeff}) holds \emph{without any
assumption on the form of the tensor $\cA_{nl}$}, aside from the
condition that no $\h$'s be present in the linear combination
(\ref{geomtens}). This means that by simply imposing the Bianchi
identity on $\cA_{nl}$ both the double-tracelessness and the
compensator form $\cA_{nl} \, = \, \cF \, - \ 3 \, \pr^{\, 3}
\a_{nl}$ emerge as direct consequences.

To these results one
should finally add the explicit solution for $\cD_{nl}$,
\begin{eqnarray}
 && {\cal D}_{nl} \, = \, \frac{1}{2} \ \sum_{k=2}^{\, n+1} \, a_k \, \left\{ \frac{1}{2\,k \, - \, 3} \, \frac{\partial^{\, 2 \, (k -
2)}}{\Box^{\, k - 2}} \ {\cal F}^{(n+1)\; [k]} \, + \, \frac{2 n + 4 k + 1}{2 (2k - 1) \, (n - k + 1)} \ \frac{\partial^{\, 2 \, (k -
1)}}{\Box^{\, k - 1}}\,
{\cal F}^{(n+1)\; [k + 1]} \right. \nonumber \\
&& \left.+ \, \frac{n +  k + 1}{2 (n - k) \, (n - k + 1)} \ \frac{\partial^{\, 2 \, k }}{\Box^{\, k}}\ {\cal F}^{(n+1)\; [k + 2]} \right\}
\end{eqnarray}
which allows to complete the construction of non-local Einstein-like tensors leading to correct current exchanges. In order to show that these
tensors are indeed divergence-free, in the odd-spin case, after all possible traces of $\cD_{nl}$ are computed and all divergences are turned
into further traces via the Bianchi identities, one can notice that the result only involves divergences of $\cF^{(n+1)\; [n]}$, that vanish
identically because of eq.~(\ref{bianchinn}).

As an example, the modified Fronsdal tensors for the cases of spin
$s=3,4$ and $5$ read
\bea \cA_3 \, &=& \, \fr{1}{\Box} \, \prd \cR^{\, \pe} \, + \,
\fr{\pr^{\, 2}}{2\,
\Box^{\, 2}}  \, \prd \cR^{\, \pe \pe} \,   ,\\
 \cA_4 \, &=& \, \fr{1}{\Box} \, \cR^{\, \pe \pe} \, + \, \12 \,
\fr{\pr^{\, 2}}{\Box^{\, 2}} \, \cR^{\, \pe \pe \pe} \, - \, 3 \,
 \fr{\pr^{\, 4}}{\Box^{\, 3}} \, \cR^{\, [4]}\, ,\\
 \cA_5 \, &=& \, \fr{1}{\Box^{\, 2}} \, \prd \cR^{\, \pe \pe} \, + \, \fr{2}{3} \,
\fr{\pr^{\, 2}}{\Box^{\, 3}} \, \prd \cR^{\, \pe \pe \pe} \, - \, 3
\, \fr{\pr^{\, 4}}{\Box^{\, 4}} \,  \prd \cR^{\, [4]} \, ,
\label{nlocleqs} \eea
while the corresponding $\cD$ tensors are similarly given by
\bea \cD_4 \, &=& \,-\fr{3}{8} \,
\fr{1}{\Box} \, \cR^{\, [4]}\, ,\\
\cD_5 \, &=& \,- \, \fr{5}{8\, \Box^{\, 2}}
\, \prd \cR^{\, [4]} \, .
\eea
In the absence of sources, taking successive traces of
the Lagrangian equation of motion coming from
(\ref{enl})
\be \cA_{nl} \, - \,  {1\over 2} \, \eta \, \cA_{nl}^{\, \pe} \, +
\, \eta^{\, 2} \,  \cB \, = \, 0 \ , \ee
one can reduce it to $\cA_{nl} \, = \, 0$. This last equation, in
turn, can be shown to imply the non-lagrangian equations of
\cite{fs1}, eqs. (\ref{oddcurv}) and (\ref{evencurv}), by making
careful use of the Bianchi identities (\ref{bianchinn}). For
example, for the spin $4$ case, once $\cB$ and $\cA_{nl}^{\, \pe}$
are shown to vanish one is left with
\be \cF^{(2)} \, + \, \12 \, \fr{\pr^{\, 2}}{\Box} \, \cF^{(2)\, \pe} \, - \, 3 \, \fr{\pr^{\, 4}}{\Box^{\, 2}} \, \cF^{(2)\, \pe \pe} \, = \, 0
\, , \ee
whose trace implies
\be \cF^{(2)\, \pe} \, - \, \fr{\pr^{\, 2}}{\Box} \, \cF^{(2)\, \pe \pe} \, = \, 0\, . \label{f2prime} \ee
Taking the divergence of this last relation, and using the identity $\prd \cF^{(2)\, \pe} \, = \, \12 \, \pr \cF^{(2)\, \pe \pe}$ it is then
possible to show that $\cF^{(2)\, \pe \pe} \, = \, 0$, which implies, via eq.~(\ref{f2prime}), that $\cF^{(2)\, \pe} \, = \, 0$ and finally
$\cF^{(2)} \, = \, 0$, eq.~(\ref{evencurv}), as previously advertised.


\scs{Conclusions}

In this paper we have examined a number of problems that present themselves when higher-spin gauge fields interact with external currents, the
key motivation for our analysis being the precise comparison between Fronsdal's constrained formulation of \cite{fronsdal} and the unconstrained
Lagrangian formulations of \cite{fs1} and \cite{fs3}. To this end, we have began by streamlining the results of \cite{fs3}, that are here
presented in terms of a few invariant structures which also extend rather
simply to the interesting cases of $AdS$ backgrounds. The subsequent
Sections have dealt with the precise comparison between the different
available formulations in the presence of external sources. These provide
an important testing ground for the Lagrangian equations,
and indeed while the local formulation of \cite{fs3} is directly equivalent to
Fronsdal's constrained formulation of \cite{fronsdal},
 this is not the case for the Lagrangian equation proposed in \cite{fs1}.
 In Section 4 we
have thus displayed the precise map between the minimal unconstrained
local Lagrangian formulation for fully symmetric tensors and
spinor-tensors presented in \cite{fs3} and the non-local formulation of \cite{fs1}, and we have proposed a new set of non-local Lagrangian
equations that lead to a proper current exchange and, in the free case, can be reduced to the non-Lagrangian equations (\ref{oddcurv}) and
(\ref{evencurv}). Let us stress that the non-local geometric form thus identified is uniquely determined by the procedures described in Section
4.

\vskip 24pt \noindent{ NOTE ADDED} \vskip 12pt \noindent The current exchange of eq.~(\ref{cexch}) and its fermionic counterpart determine the
van Dam-Veltman-Zakharov discontinuity \cite{vdvz} in flat space for this whole classes of (bosonic and fermionic) higher-spin fields. These
expressions depend on the dimension $D$ of space time, and the discontinuity follows directly from the comparison of the $D$-dimensional result
with the corresponding one in $D+1$ dimensions. This is the case since the massless theory in $D+1$ dimensions, after a suitable reduction \`a
la Scherk-Schwarz \cite{ss}, describes irreducibly a massive field in the Stueckelberg mode. The extension of this result to an $AdS$
background, along the lines of what was done by Higuchi and Porrati for $s=2$ \cite{porrati}, is quite interesting and will be discussed
elsewhere.


\vskip 24pt


\section*{Acknowledgments}


We are grateful to the CPhT of the Ecole Polytechnique, to the
APC-Paris VII, to the LPT-Orsay, to the Universit\`a Roma Tre and to
the CERN Theory Division, where the stay of A.S., while most of this
work was being done, was supported in part by the Scientific
Associates Program, for the kind hospitality extended to us. D.F. is
also grateful to the MPI-Albert Einstein Insitute, where part of
this work was done. The present research was also supported in part
by INFN, by the MIUR-PRIN contract 2003-023852, by the EU contracts
MRTN-CT-2004-503369 and MRTN-CT-2004-512194, by the INTAS contract
03-51-6346, and by the NATO grant PST.CLG.978785.

\newpage

\begin{appendix}


\scs{Notation and conventions}\label{app:SYM}


We use the ``mostly positive'' convention for the space-time signature, and typically omit symmetrized indices in tensor relations. Hence, our
gamma matrices satisfy ${\g^{\, 0}}^{\, \dagger} = - \,\g^{\, 0}$, \, ${\g^{\, i}}^{\, \dagger} = + \,\g^{\, i}$,\, $\g^{\, 0} {\g^{\, \m}}^{\,
\dagger} \g^{\, 0}  = \g^{\, \m}$. In addition, a ``prime'' always denotes a trace: $U^{\; \prime}$ is thus the trace of $U$, while $U^{\;
\prime\prime}$ is its double trace. A generic multiple trace, however, is denoted by a bracketed suffix, so that $U^{[n]}$ is the $n$-th trace
of $U$. This notation is a very convenient method of streamlining the presentation, but it also results in an effective calculational procedure.
This is especially true in flat space, but also in $AdS$ backgrounds provided some care is clearly exercised with the resulting non-commuting
covariant derivatives. To take full advantage of the compact notation, one needs to make repeated use of a number of identities, which reflect
some simple combinatorics. These rest on our convention of working with symmetrized objects \emph{not} of unit strength, which is convenient in
this context but is not commonly used. For instance, given of vectors $A_\mu$ and $B_\nu$, $A\, B$ here stands for $A_\mu\, B_\nu + A_\nu\,
B_\mu$, without additional factors of two. The key identities are then:
\begin{eqnarray}
  \left( \pr^{\, p} \, \vf  \right)^{\, \pe} & = & \ \Box \,
  \pr^{\, p-2} \, \vf \ + \, 2 \, \pr^{\, p-1} \,  \prd \vf \ + \, \pr^{\, p} \,
\vf^{\, \pe} \,  , \nonumber \\
 \partial^{\, p} \, \partial^{\, q} & = & \ {p+q \choose p} 
\partial^{\, p+q} \ ,
\nonumber \\
 \partial \cdot  \left( \partial^{\, p} \ \vf \right) \ & = & \ \Box \
\partial^{\, p-1} \ \vf \ + \
\partial^{\, p} \ \partial \cdot \vf \ ,  \\
 \partial \cdot  \eta^{\, k} & = & \ \partial \, \eta^{\, k-1} \ , \nonumber\\
 \left( \eta^k \, \vf  \,  \right)^{\, \prime} & = & \ \left[ \, D
\, + \, 2\, (s+k-1) \,  \right]\, \eta^{\, k-1} \, \vf \ + \ \eta^k
\, \vf^{\, \prime} \ , \nonumber \\
(U \ V)^{\; \prime} & =& U^{\; \prime} \ V + U \ V^{\; \prime} + 2
\, U \cdot V \ , \nonumber \\
 \eta \ \eta^{n-1} & = & n \, \eta^n \ .\nonumber
 \label{etak}
\end{eqnarray}
As anticipated, the basic ingredient in these expressions is the
combinatorics, which is simply determined by number of relevant
types of terms on the two sides. Thus, for a pair of flat
derivatives, $\partial \,
\partial = 2 \, \partial^{\, 2}$ reflects the fact that, as a result of
their commuting nature, the usual symmetrization is redundant
precisely by the overall factor of two that would follow from the
second relation. In a similar fashion, for instance, the last
identity reflects the different numbers of terms generated by the
naive total symmetrization of the two sides: $\left( {2 n} \atop {2}
\right) \times (2 n -1)!!$ for the expression on the \emph{l.h.s},
and $(2 n + 1)!!$ for the expression on the \emph{r.h.s.}.
\end{appendix}

\newpage


\end{document}